\documentclass[article]{article}
\usepackage{graphicx}
\usepackage[utf8]{inputenc}
\usepackage[margin=1in]{geometry}
\usepackage{authblk}
\usepackage{tabularx}
\usepackage{hyperref}
\usepackage{natbib}
\usepackage[export]{adjustbox}
\usepackage{array,calc}
\usepackage{color}
\usepackage{sidecap}
\usepackage{floatrow}
\usepackage{subfig}

\begin{document}

\title{Data-driven biological network alignment that uses  topological, sequence, and  functional information}
\author[ ]{Shawn Gu}
\author[*]{Tijana Milenkovi\'c}
\affil[ ]{Department of Computer Science and Engineering, Eck Institute for Global Health, and Center for Network and Data Science, University of Notre Dame, Notre Dame, IN 46556}
\affil[*]{\normalsize{To whom correspondence should be addressed (email: tmilenko@nd.edu)}}
\date{}
\maketitle

\abstract{Network alignment (NA) can transfer functional knowledge between species' conserved biological network regions.  Traditional  NA assumes  that  it  is topological similarity (isomorphic-like matching) between network regions that  corresponds  to  the  regions'  functional  relatedness.  However, we recently found  that functionally unrelated proteins are as topologically  similar  as  functionally  related  proteins. So, we redefined NA as a data-driven framework, TARA, which learns from network and protein functional data what kind of topological \emph{relatedness} (rather than similarity)  between proteins corresponds  to  their  functional  relatedness. TARA used topological information (within each network) but not sequence information (between proteins across networks). Yet, TARA yielded higher protein functional prediction accuracy than existing NA methods, even those that used both topological and sequence information. Here, we propose TARA++ that is also data-driven, like TARA and unlike other existing methods, but that uses across-network sequence information on top of within-network topological information, unlike TARA. To deal with the within-and-across-network  analysis, we adapt social network embedding to the problem of biological NA. TARA++  outperforms protein functional prediction accuracy of existing methods.}

\maketitle


\section{Introduction} \label{sec:intro}

\subsection{Background}


Many proteins remain functionally unannotated \citep{ellens2017confronting}. A popular way to uncover missing annotations is to transfer functional knowledge across proteins of different species. This task of \emph{across-species} protein functional prediction is the focus of this paper. The orthogonal task of \emph{within-a-species} protein functional prediction, where a function of a protein in a species is predicted from function(s) of other protein(s) in \emph{the same} species
\citep{shehu2016survey}, 
 is out of the scope.

Genomic sequence alignment is commonly used for the task of across-species protein functional prediction, by transferring functional knowledge between conserved (aligned) sequence regions of proteins in different species. However, sequence alignment often fails: many sequence-similar proteins do not perform the same function(s), i.e.,  are functionally unrelated, and many sequence-dissimilar proteins are functionally related \citep{gu2019data}; ``functionally related'' means that, according to current Gene Ontology (GO) annotation data \citep{ashburner2000gene}, two proteins share a GO term, while ``functionally unrelated'' means that they share no GO terms. For example, of all yeast-human sequence orthologs from YeastMine \citep{balakrishnan2012yeastmine},  $\sim$42\% are \emph{not} functionally related \citep{gu2019data}. Such imperfect performance of sequence  alignment could be due to sequence  alignment failing to  consider interactions between the genes, i.e., their protein products. However, a protein does not function alone. Instead, it interacts in complex networked ways with other proteins. So, accounting for protein-protein interactions (PPIs) is important for better protein functional prediction across species. 



Luckily, large amounts of PPI network data are available \citep{chatr2017biogrid}. Hence, network alignment (NA)  can be used to compare PPI networks of different species, in order to find a ``good'' mapping between their nodes (proteins), i.e., a node mapping that uncovers regions of high network topological (and often sequence) conservation between the species; conservation typically means similarity. So, analogous to sequence alignment, NA can be used to transfer functional knowledge between conserved (aligned) PPI network, rather than just sequence, regions of different species \citep{faisal2015post, meng2016local, emmert2016fifty, elmsallati2016global, guzzi2017survey}.
While we focus on computational biology, NA is also applicable to many other domains \citep{emmert2016fifty}.

NA can be categorized into several broad types, whose high-level input/output/goal differences are as follows (more detailed algorithmic differences between specific NA methods are discussed in Section \ref{sec:rel-work}). 

First, NA can be pairwise (aligns two networks) or multiple (aligns three or more networks) \citep{faisal2015post, guzzi2017survey}. We focus on  pairwise NA because current multiple NA is more computationally complex  \citep{vijayan2018multiple} while also generally less accurate \citep{vijayan2019pairwise} than current pairwise NA. 

Second, NA can be local or global \citep{meng2016local, guzzi2017survey}, like sequence alignment.
Local NA aims to find highly conserved network regions but usually results in such regions being small. Global NA aims to maximize overall network similarity; while it usually results in large aligned network regions, these regions are suboptimally conserved. Both have their own (dis)advantages \citep{meng2016local, guzzi2017survey}. (More on local versus global NA follows shortly. We first need to define  one-to-one versus many-to-many NA.)

Third, NA can be one-to-one (each node can be aligned to exactly one distinct node in another network) or many-to-many (a node may be aligned to more than one node in another network). 

Traditionally, given networks $G_1(V_1,E_1)$ and $G_2(V_2,E_2)$, local NA has meant the same as many-to-many NA:  a relation $R \subseteq V_1 \times V_2$. Also, global NA has meant the same as one-to-one NA: an injective function $f: V_1 \rightarrow V_2$. Over time, local one-to-one  and global many-to-many NA methods have been proposed. So, both local and global NA are now $R \subseteq V_1 \times V_2$. The two differ in how many nodes are covered by the aligned node pairs in $R$ -- much fewer for local than global NA.  

As global NA has received more attention recently than local NA, we focus on global NA.
Both one-to-one and many-to-many alignments can be used in our considered task of across-species protein functional prediction. 
Yet, it is many-to-many NA methods that are the state-of-the-art in this task, which is why our considered methods happen to be many-to-many.

Fourth, three NA method groups exist based on how input data are processed. \emph{The first group} consists of NA methods that, given two PPI networks, calculate each node's feature  using only the topological information within the given node's own network. 
As such, we refer to them as  \emph{within-network-only} NA methods. The nodes' topological features, which aim to summarize the nodes' extended PPI network neighborhoods,  are then used in various  alignment processes (Section \ref{sec:rel-work}). For state-of-the-art NA methods from this group, the topological features are based on graphlets \citep{milenkovic2008uncovering}, which are subgraphs, i.e., small, Lego-like building blocks of networks.  
\emph{The second group} consists of NA methods that, given  two PPI networks and also sequence information for nodes across networks, first calculate each node's topological feature in the same way as within-network-only methods, and \emph{only afterwards} combine the sequence information with the topological features. 
Then, the combined data are used in various  alignment processes (Section \ref{sec:rel-work}). Because both within-network topological and across-network sequence information are used, but the two are initially processed in isolation from each other and are combined only after the fact, we refer to this second group as \emph{isolated-within-and-across-network} methods. 
Within-network-only methods can easily be used as isolated-within-and-across-network methods when sequence information is available; the latter usually lead to better alignments than the former \citep{meng2016local}.
\emph{The third group} consists of NA methods that, given two PPI networks and sequence information for nodes across networks,  first ``integrate'' the two networks into one by adding across-network ``anchor'' links (edges) between the highly sequence-similar proteins and \emph{only then} proceed with any feature extraction or alignment process. So, the third group uses both within-network topological and across-network sequence information. But, they first integrate the two data types and only then process them. 
As such, we refer to them as \emph{integrated-within-and-across-network} methods. 


\subsection{Motivation}


Regardless of which NA category they belong to, almost all existing  NA methods assume that it is topological similarity between nodes (i.e., a high level of isomorphism-like matching between their extended PPI network neighborhoods as captured by the nodes' topological features) that 
corresponds to the nodes' functional relatedness, and thus they try to align topologically similar \textcolor{black}{nodes. 
However, multiple studies observed that} while existing NA methods yield high topological alignment quality (many edges are conserved, i.e., the aligned network regions indeed have a high level of isomorphism-like match), their functional alignment quality is far from perfect (often, the aligned nodes are \emph{not} functionally related)  \citep{elmsallati2016global, meng2016local, guzzi2017survey}. 

Only recently, an attempt was made to understand this observation, and this was done by us \citep{gu2019data}. Namely, we questioned the key assumption of current NA -- that topologically similar nodes correspond to functionally related nodes. We found for both synthetic and PPI networks  that the functionally related nodes were only marginally more topologically similar than the functionally unrelated nodes, no matter which  topological similarity measure was used \citep{gu2019data}. 

\begin{figure}[ht!]
 \centering    
        \sidesubfloat[]{\includegraphics[width=0.25\textwidth]{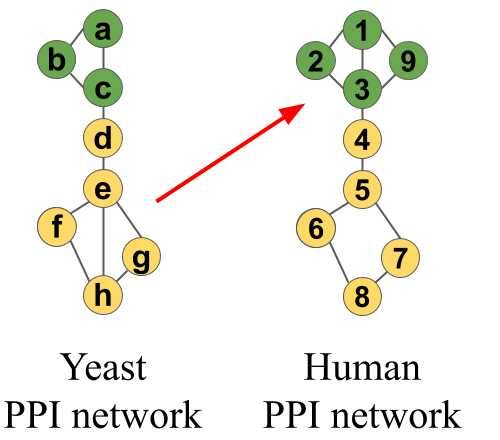}}
     \sidesubfloat[]{\includegraphics[width=0.25\textwidth]{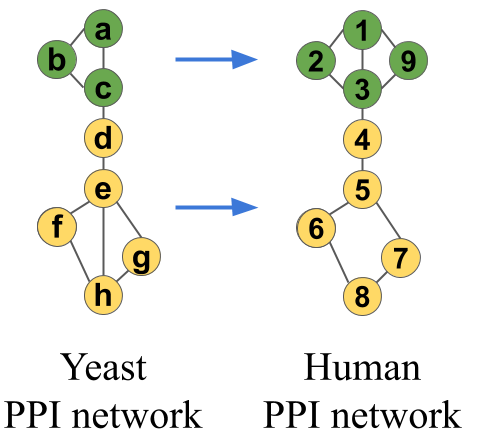}}
 \caption{\label{fig:sim-vs-rel} Illustration of topological \textbf{(a)}  similarity vs. \textbf{(b)} relatedness.}
\end{figure}

This shocking result - the current NA assumption failing -- led us to redefine the NA problem as a data-driven framework, which learns from PPI network and protein functional data what kind of ``topological relatedness'' between proteins corresponds to the proteins' functional relatedness, without assuming that topological relatedness  means topological similarity. 
%
To understand this framework, we next explain topological similarity versus topological relatedness (Fig. \ref{fig:sim-vs-rel}).

Suppose that: (i) PPI networks of yeast and human are being aligned, (ii) the  toy networks in Fig. \ref{fig:sim-vs-rel} are parts of the full networks, (iii) each node  performs either the ``green'' or ``yellow'' function, and (iv) because of \textcolor{black}{incompleteness/noisiness of PPI network data or} molecular evolutionary events such as gene duplication, deletion, or mutation, the green functional module in human (nodes $1$, $2$, $3$, and $9$) has an extra protein compared to the green module  in yeast (nodes $a$, $b$, and $c$), and the yellow module  in yeast has an extra  interaction compared to the yellow  module  in human (Fig. \ref{fig:sim-vs-rel}). 
An NA method based on topological similarity will  align yellow nodes $e$, $f$, $g$, and $h$ in yeast to green nodes $1$, $2$, $3$, and $9$ in human (Fig. \ref{fig:sim-vs-rel}(a)), because both node sets form the same subgraph -- a square with a diagonal, i.e., because the set of yellow nodes in yeast are topologically more similar to the set of green nodes in human than to the set of yellow nodes in human. However, this alignment is functionally incorrect because yellow and green nodes perform different functions. Instead, our NA framework based on topological relatedness will use the topological and functional data to learn that a triangle in yeast ($a$, $b$, and $c$) should be aligned to a square-with-diagonal in human ($1$, $2$, $3$, and $9$) because both perform the same function (green), and that a square-with-diagonal in yeast ($e$, $f$, $g$, and $h$) should be aligned to a square in human ($5$, $6$, $7$, and $8$) because both perform the same function (yellow) (Fig. \ref{fig:sim-vs-rel}(b)). Then, in other parts of the networks, our framework will try to align these learned patterns, to transfer knowledge between them. 
Loosely speaking, topological relatedness aims to account for \textcolor{black}{data noisiness/incompleteness,} evolutionary events, \textcolor{black}{or other, yet-to-be-discovered factors} that are likely to break the isomorphism-like assumption of the traditional topological similarity-based NA.


We named our topological relatedness-based NA framework TARA \citep{gu2019data}. TARA uses supervised classification to learn what topological patterns should be aligned to each other. Given (i)  a set of node pairs  across the networks being aligned, 
such that  the \textcolor{black}{nodes} in a given pair are functionally related,
(ii) a set of 
node pairs across the networks such that nodes in a given pair are functionally unrelated,
and (iii) graphlet-based network topological features of each node pair, TARA divides the node pairs
into training and testing data. Then, it uses a classifier to learn from the training data what graphlet features  distinguish between the functionally related and functionally unrelated node pairs.  Next, given node pairs from the testing data and their graphlet features, TARA predicts whether the nodes in a given pair are functionally related or unrelated. Node pairs predicted as functionally related are added to TARA's alignment, and this alignment is given to an established across-species protein functional prediction methodology \citep{meng2016local} to obtain a list of protein functional annotations (i.e., protein-GO term pairs). 



By learning topological relatedness patterns, TARA outperformed, in the task of across-species protein functional prediction between yeast and human, three state-of-the-art NA methods, WAVE \citep{sun2015simultaneous}, SANA \citep{mamano2016sana}, and PrimAlign \citep{kalecky2018primalign}.
To better understand the implications of these results, it is important to understand what type, i.e., within-network-only, isolated-within-and-across-network, or integrated-within-and-across-network, each method is.
%
TARA, WAVE, and SANA are all within-network-only methods. They also all use graphlet-based topological node features. Their key difference is that TARA \textcolor{black}{is supervised, ie., it uses topological relatedness, while WAVE and SANA are unsupervised, i.e., they use topological similarity.} Thus, WAVE and SANA were the most fairly comparable methods to TARA. So, we could fairly evaluate whether moving from WAVE's and SANA's  topological similarity to TARA's supervision-based topological relatedness helped. TARA significantly outperformed WAVE and SANA, so we could conclude that it did help.  
%
PrimAlign is one of very few existing integrated-within-and-across-network methods. Because PrimAlign was already shown to outperform many isolated-within-and-across-network methods \citep{kalecky2018primalign} on the \emph{exact same data} as in TARA's evaluation \citep{gu2019data}, there was no need to evaluate TARA against any  methods of that type. 
Importantly, TARA still outperformed PrimAlign, despite the former not using any sequence information. This  already showed how powerful the supervised NA paradigm is. 
In this study, we push the boundary further. TARA ``only'' showed that going from unsupervised to supervised for within-network-only methods improved alignment accuracy, but we also already know that going from within-network-only to isolated-within-and-across-network to integrated-within-and-across-network in the unsupervised context also improves accuracy. So, a method that is both supervised and of the integrated-within-and-across-network type should further improve alignment accuracy. Thus, here, we propose the first ever method of this type.

\subsection{Our contributions}


We introduce TARA-TS (\underline{TARA} within-network \underline{T}opology and across-network \underline{S}equence information) as a novel framework implementing the above idea. Like TARA, TARA-TS is supervised. 
Unlike TARA and like PrimAlign, TARA-TS extracts features from an integrated yeast-human network. 
\textcolor{black}{As a solution to feature extraction, we  leverage} the extensive research on graph representation learning \citep{cai2018comprehensive}, 
which embeds nodes of a network into a low dimensional space such that network structure is preserved; the low-dimensional node representations are then used as node features. Network embedding has primarily been studied in the social network domain. So, we extend it to the domain of computational biology: TARA-TS generalizes a prominent network embedding method that was proposed for single-network analysis in machine learning tasks such as node classification, clustering, and link prediction, to the multi-network task of biological NA. 
%
Given the node features extracted by network embedding, TARA-TS works just as TARA to produce an alignment. Then, we use this alignment for across-species protein functional prediction. 

We compare prediction accuracy of TARA-TS (pairwise, global, many-to-many, integrated-within-and-across-network, supervised) with accuracies of TARA and PrimAlign, as they are state-of-the-art NA methods that were already shown to outperform many other existing NA methods on the \emph{exact same data} as what we use here. So, by transitivity, if TARA-TS is shown to be superior to TARA and PrimAlign, this will mean that TARA-TS is superior to the other existing methods as well. Also, of all existing methods, TARA and PrimAlign  are the most similar and thus fairly comparable to TARA-TS. Namely, TARA is pairwise, global, many-to-many, and supervised, like TARA-TS. The  difference is that TARA is a within-network-only method while TARA-TS is an integrated-within-and-across-network method. So, we can fairly test the effect of going from within-network-only to integrated-within-and-across-network in the supervised context. PrimAlign is a pairwise, global, many-to-many, and integrated-within-and-across-network method, like TARA-TS. The difference is that PrimAlign is unsupervised while TARA-TS is supervised. So, we can fairly test the effect of going from unsupervised to supervised for integrated-within-and-across-network methods.

When we compare TARA-TS against TARA, we actually compare whether using across-network sequence information on top of  within-network topological information leads to more accurate predictions, as we expect. Shockingly, we find that TARA-TS and TARA are almost equally as accurate. Closer examination reveals that their quantitatively similar results are \emph{not} because the two methods are predicting the same information (which would make one of them redundant). Instead, their predicted protein functional annotations are quite complementary. So, we then look at those predictions (protein-GO term associations) that are made by both methods, only those predictions made by TARA-TS but not TARA, and only those predictions made by TARA but not TARA-TS. We find the former (the overlapping predictions) to be more accurate than the predictions made by any one of TARA-TS or TARA alone. So, we take this overlapping version of TARA-TS and TARA as our final method, TARA++. In a sense, TARA++ is integrating state-of-the-art research knowledge across computational biology and social network domains, by combining TARA's  graphlet-based topology-only features with TARA-TS's embedding-based topology-and-sequence features, each of which boosts the other's performance. Very few studies have explored such a promising direction to date \citep{nelson2019embed}.
%
%
%
Importantly, we find that TARA++ not only outperforms TARA but also PrimAlign.
%
%

\subsection{Related work} \label{sec:rel-work}

First, we discuss  within-network-only and isolated-within-and-across-network methods. They have two  parts. First, similarities are computed for all pairs of nodes across networks. For within-network-only methods, these are  topological similarities (computed by comparing the nodes' topological features). For isolated-within-and-across-network methods,  these are a weighted sum of the nodes' topological   and sequence  similarities. Second, an alignment strategy aims to maximize the total similarity over all aligned nodes while also conserving many edges. 
Two types of alignment strategies exist. One type is ``seed-and-extend'', which progressively builds an alignment by adding to it one node pair at a time.
WAVE \citep{sun2015simultaneous}, when paired with graphlet-based topological similarities,  is a state-of-the-art  method of this type. 
The other type is a ``search algorithm'' that optimizes an objective function over the solution space of possible alignments.
We pioneered search algorithm-based NA with  MAGNA and MAGNA++  \citep{saraph2014magna,MAGNAPP}. The more recent SANA \citep{mamano2016sana} is a state-of-the-art approach of this type, whose
objective function is generally graphlet-based.

Next, we discuss integrated-within-and-across-network NA methods. PrimAlign \citep{kalecky2018primalign} is a state-of-the-art method of this type. After linking  networks being aligned via anchors, PrimAlign creates a Markov chain out of the integrated network, converting the edge weights to transition probabilities (in an unweighted network, the weights are set to 1 before converting to transition probabilities). The chain is then repeatedly transitioned until convergence, redistributing the across-network node pair scores using a PageRank-like algorithm. Node pairs across networks that are above some threshold are then taken as the alignment. 

MUNK also links the original networks via  anchors, but it uses  matrix factorization 
to obtain an alignment \citep{fan2019functional}. In our preliminary analyses, MUNK's similarity scores were not able to distinguish between functionally related and functionally unrelated proteins. Furthermore,  Nelson et al. \citep{nelson2019embed} found IsoRank \citep{singh2008global} to outperform MUNK, despite the former being an early method and the latter a recent method. IsoRank was already outperformed by many NA methods that appeared after it, which in turn were outperformed by WAVE and SANA, which were then outperformed by TARA and PrimAlign (see below). Because we consider TARA and PrimAlign in this study, there is no need to also consider MUNK, an inferior method.

Unlike TARA++, the above methods do not use functional (GO) information to produce alignments but only to evaluate them. DualAligner \citep{seah2014dualaligner} does use such information, but not to determine classification labels (``functionally related'' and ``functionally unrelated'') like TARA++ does. Instead, the method aligns groups of nodes that are all annotated with a given GO term, and then seeds-and-extends around these groups to match proteins that do not have any GO annotations, resulting in the final alignment. We do not consider DualAligner in this study, as it is quite old. More recent, state-of-the-art methods have appeared since. 

The above methods are unsupervised. Many other such methods exist \citep{guzzi2017survey}. TARA and PrimAlign, which we consider in this study, already outperformed the other methods, including AlignMCL, AlignNemo, CUFID, HubAlign, IsoRankN, L-GRAAL, MAGNA, MAGNA++, MI-GRAAL, NETAL, NetCoffee, NetworkBLAST, PINALOG, SANA, SMETANA, and WAVE \citep{sun2015simultaneous, mamano2016sana, kalecky2018primalign, gu2019data}. In turn, these outperformed GHOST, IsoRank, NATALIE, PISwap, and SPINAL \citep{mamano2016sana}. This, plus TARA and PrimAlign being the most similar and thus fairly comparable to TARA++, is why we focus on these two existing methods. 
Also, some supervised methods (besides TARA, already discussed) exist, as follows.

\textcolor{black}{IMAP \citep{cao2017imap} uses supervised learning  differently than TARA++. As input, IMAP requires a starting (unsupervised, topological similarity-based) alignment between two networks; as such, it still suffers from the topological similarity assumption. Then, it obtains graphlet features for node pairs. Node pairs from the starting alignment form the positive class, while the other node pairs are sampled to form the negative class. Then, IMAP trains a linear regression classifier on these two classes. After, this data is ``re-classified'', but instead of assigning a class, IMAP assigns a score corresponding to the probability that the two nodes should be aligned. A matching algorithm (e.g., Hungarian) is applied to these scores to form a new alignment, which is then fed back to IMAP. This process iterates while alignment quality improves.  We did try to test IMAP. Its code was not available. Our attempts at implementing IMAP ourselves led to significantly worse results than those reported in the IMAP paper. So, we could not consider IMAP in our evaluation.}

MEgo2Vec \citep{zhang2018mego2vec}, also supervised, is a social NA method for matching user profiles across different online media platforms. Features of user profiles are obtained using graph neural networks and natural language processing techniques, and these are used to train a classifier to predict whether two profiles from different platforms correspond to the same person. A big part of MEgo2Vec is the various text processing techniques to match users' names, affiliations, or research interests, meaning that it cannot  be easily applied to PPI networks.


\section{Methods}



\subsection{Data} \label{sec:methods-data}


As typically done in NA studies, we analyze PPI networks of yeast (5,926 nodes and 88,779 edges) and human (15,848 nodes and 269,120 edges) from BioGRID \citep{chatr2017biogrid}. Also, like  the PrimAlign study \citep{kalecky2018primalign},  we consider 55,594 yeast-human protein pairs with E-value sequence similarities $\leq 10^{-7}$ as anchor links.

Our supervised NA framework requires knowledge about whether two proteins are functionally related. As typically done, we define functional relatedness using  GO annotation data (from August 2019). Considering biological process GO terms and experimentally inferred protein-GO term annotations (evidence codes EXP, IDA, IPI, IMP, IGI, or IEP), if at least $k$ GO terms  are shared between a yeast protein and a human protein, we define that protein pair as functionally related; we vary $k$ from 1 to 3. Regardless of the $k$ value, we define a protein pair as functionally unrelated if the two proteins share no GO terms \emph{of any kind}. This gives three ground truth datasets:  \textit{atleast1-EXP}, \textit{atleast2-EXP}, and \textit{atleast3-EXP}. 

While traditionally every GO term available in a given ground truth dataset has been  considered for NA evaluation, recent work \citep{hayes2017sana} suggested that accounting for the frequency of GO terms (how many proteins a GO term annotates out of all proteins analyzed)  is important for reducing GO term redundancy. Indeed, in our TARA study, we found that considering rarer GO terms led to higher protein functional prediction accuracy \citep{gu2019data}. So, here, we consider the same three GO term rarity thresholds as in the TARA study: (i) all GO terms (i.e., ALL), which corresponds to traditional NA evaluation, (ii) GO terms that appear 50 times or fewer (i.e., threshold of 50), and (iii) GO terms that appear 25 times or fewer (i.e., threshold of 25). 

For a given GO term rarity threshold, all GO terms not satisfying the threshold are filtered out. Then, for each atleast$k$-EXP ground truth dataset, only proteins that share at least $k$ GO terms from the remaining list are considered to be functionally related, and still, proteins that share no GO terms, regardless of rarity, are considered to be functionally unrelated. For example, proteins that share at least one (experimentally inferred biological process) GO term, such that each GO term annotates 50 or fewer proteins, are considered functionally related in the ``atleast1-EXP at the 50 GO term rarity threshold'' dataset. There is a total of nine such ``ground truth-rarity'' datasets, resulting from combinations of the three atleast$k$-EXP ground truth datasets and the three GO term rarity thresholds. 


\subsection{TARA-TS's feature extraction methodology} \label{sec:tara-to-ts}


\textcolor{black}{TARA-TS needs to extract features that capture both within-network topological and across-network sequence information from the integrated network, which consists of 21,774 nodes (5,926 yeast + 15,848 human proteins) and 413,493 edges (88,779 yeast PPIs + 269,120 human PPIs + 55,594  anchor links). We examine several feature extraction approaches.}

\textcolor{black}{First, we use the same feature extraction method as TARA, simply applied to the integrated network rather than the two individual networks. TARA relies on graphlets, small subgraphs (a path, triangle, square, etc., generally up to five nodes). Graphlets are often used to summarize the extended neighborhood of a node into its feature vector, as follows. For a node, for each automorphism orbit (intuitively, node symmetry group) in a graphlet, one can count how many times the node touches each graphlet at each of its orbits. The resulting counts for all considered graphlets/orbits are the node's \textit{graphlet degree vector} (GDV) \citep{milenkovic2008uncovering}. Then, to obtain the feature of a node pair, TARA takes the element-wise absolute difference of the nodes' GDVs. So TARA-TS can apply the same graphlet counting procedure to the integrated network, obtaining the GDV for each yeast and human node, and taking the absolute difference of two nodes' GDVs to obtain the feature vector of the yeast-human node \emph{pair}. We refer to this version of TARA-TS as ``TARA-TS (graphlets)''. }

\textcolor{black}{Second, we use a prominent network embedding method on the integrated network to extract features, namely node2vec \citep{grover2016node2vec}. This method uses random walks to explore the neighborhood of a node in a network. For a node $u$, random walks starting at $u$ are performed, and the sequence of nodes visited by each random walk is recorded. The number of random walks performed per node is controlled by the parameter ``Number of walks per source (\texttt{-r:})'', and the length of a random walk is controlled by the parameter ``Length of walk per source (\texttt{-l:})''. This process is repeated for every node in the network. Then, a skip-gram model is applied over all sequences of nodes and the feature vector of each node is obtained. The only way to use node2vec, a single-network method, in the multi-network NA task, is to first integrate the two networks via anchor links, as we do. Otherwise,  node2vec fails if applied to the two networks individually \citep{gu2018graphlets}. We first apply node2vec to the integrated network with the default parameters to obtain a feature vector for each \emph{node}. Then, as suggested by the node2vec study \citep{grover2016node2vec}, to get the feature vector of a yeast-human node \emph{pair}, we take the element-wise average of the nodes' feature vectors. We refer to this version of TARA-TS as ``TARA-TS (node2vec)''.}

\textcolor{black}{We use node2vec over other network embedding methods for three reasons. (i) Even more recent  methods, when evaluated in their own papers, achieve similar performance as node2vec in many tasks. So, we do not expect them to outperform node2vec in our task. (ii) The node2vec source code is available and well documented, unlike for many other methods. (iii) The goal of this study is not to find the absolute best feature vector for supervised NA, but to test how combining topological and sequence information in supervised NA affects protein functional prediction. If using node2vec already improves upon current NA methods, then using any more sophisticated ways to extract features will only improve further. In our proposed framework, features from any new extraction method can simply be ``swapped'' in, allowing flexibility for further advancements.}

\textcolor{black}{Third, node2vec does not capture heterogeneous information in the integrated network, i.e., does not distinguish between different types of nodes (yeast and human) or edges (yeast PPIs, human PPIs, and yeast-human sequence-based anchor links). So, we also test metapath2vec \citep{dong2017metapath2vec}, i.e., node2vec generalized to heterogeneous networks. Metapath2vec requires the user to define ``metapaths'', which direct how the random walks move. A metapath example is  ``human-human-yeast-yeast'' (or ``human$\times$2 $\rightarrow$ yeast$\times$2''): start at a human node, move to a randomly chosen neighboring (RCN) human node, move to an RCN yeast node, and move to  an RCN yeast node. This metapath is extended such that its length is as close as possible to the \texttt{-l:} parameter value. For example, if this value is 12, then this metapath would be repeated $\left \lfloor{12/4}\right \rfloor  = 3$ times. Then, given a node $u$ and the extended metapath, random walks starting at $u$ are performed such that the nodes visited follow the constraints of the metapath, and the sequence of nodes visited by each random walk is recorded. In the process, node2vec's \texttt{-l:} and \texttt{-r:} parameters apply to metapath2vec as well. The procedure is repeated for every node in the network. Then, a skip-gram model is applied over all sequences of nodes to obtain node features. We use the metapath2vec++ implementation of metapath2vec \citep{dong2017metapath2vec}  with the default parameters to obtain each node's feature vector, and again take the element-wise average of two nodes' feature vectors to compute the feature vector of a node \emph{pair}. Choosing ``optimal'' metapaths  is non-trivial and often the selection process involves using the same paths as those of previous studies \textcolor{black}{\citep{shang2016meta, dong2017metapath2vec}}. However, to our knowledge, this study is the first to investigate metapaths on an integrated across-species biological network. Thus, our only option is to do our due diligence and examine reasonable metapaths, to give a fighting chance to metapath2vec. 
We test these metapaths: ``human$\times$n $\rightarrow$ yeast$\times$n'' and ``yeast$\times$n $\rightarrow$ human$\times$n'' for $3 \leq n \leq 10$, ``human$\times$25 $\rightarrow$ yeast$\times$25'' and ``yeast$\times$25 $\rightarrow$ human$\times$25'', ``human$\times$50 $\rightarrow$ yeast$\times$50'' and ``yeast$\times$50 $\rightarrow$ human$\times$50'', and the combination of all of the individual metapaths (i.e., we apply the skip-gram model to all node sequences obtained from all considered metapaths). We verified that the choice of metapath does not impact alignment accuracy (results not shown). So, for simplicity, we continue with the combination of the metapaths. We refer to this version as ``TARA-TS (metapath2vec)''.}

\textcolor{black}{Henceforth, we refer to TARA-TS (graphlets), TARA-TS (node2vec), and TARA-TS (metapath2vec) as different ``TARA-TS versions''. If we just say ``TARA-TS'', the discussion applies to all three versions.}

\textcolor{black}{In theory, the heterogeneous information  could be captured not just via metapaths but also via heterogeneous graphlets \citep{gu2018homogeneous} (versus homogeneous graphlets discussed thus far). However, in practice, heterogeneous graphlet counting is infeasible for as large networks as studied in this paper, due to its exponential computational complexity. This is not an issue for homogeneous graphlet counting because methods such as Orca \citep{hovcevar2014combinatorial} rely on combinatorics to infer the counts of some (larger) graphlets from the counts of other (smaller) graphlets, significantly reducing the computational complexity.
\textcolor{black}{However, no publicly available implementation of combinatorial relationships for counting heterogeneous graphlets exists.} 
Similar holds for a method that \emph{directly} extracts the feature vector of a \emph{node pair}  \citep{hulovatyy2014revealing}, versus extracting graphlet features of \emph{individual nodes} and then combining these, as TARA does: \textcolor{black}{no combinatorial approach for direct node pair graphlets exists}. Instead, current heterogeneous and node pair graphlet counting require exhaustive graphlet enumeration and are thus infeasible.}

\textcolor{black}{Lastly, we discuss why we do not use feature vectors from PrimAlign, the next most comparable method to TARA \citep{gu2019data} that already integrates within-network topological and across-network sequence information. This is because PrimAlign's algorithmic design does not allow for feature vector extraction. Recall from Section \ref{sec:rel-work} that PrimAlign models the  integrated network as a Markov chain, which is then repeatedly transitioned until convergence. This means that the weights between every node pair are updated at the same time, based on the weights of every node pair from the previous state of the chain. So, PrimAlign operates on every node pair \emph{at once} with respect to their weights, rather than on \emph{individual nodes or node pairs} with respect to any kind of feature vector, meaning that we cannot extract such information.} 

\subsection{TARA-TS's classification and alignment generation} \label{sec:methods-supNA}



\textcolor{black}{We must first evaluate whether TARA-TS can correctly predict nodes as functionally (un)related. If not, there would be no point to use it to form an alignment. To evaluate this, we train and test a classifier as follows.}

For a given ground truth-rarity dataset (Section \ref{sec:methods-data}), the positive class consists of functionally related node pairs, and the negative class consists of functionally unrelated node pairs. Because the latter is much larger, we create a balanced dataset by undersampling the negative class to match the size of the positive class, as typically done \citep{sun2009classification}. \textcolor{black}{Due to randomness in sampling, we create 10 balanced datasets and repeat the classification process for each, averaging results over them.} 

\textcolor{black}{For a given balanced dataset, we split it into two sets: $y$ percent of the data is randomly sampled and put into one set, and the remaining $(100-y)$ percent is put into the other set. This sampling is done with the constraint that in each of the two sets, 50\% of the data instances have the positive class and 50\% have the negative class. 
Again, due to randomness in sampling, we repeat this 10 times to create 10 data splits of $y/(100-y)$ percent and repeat the classification process for each, averaging results over them.} 

\textcolor{black}{For a given $y/(100-y)$ split, we train a logistic regression classifier on the set containing $y$ percent of the data (i.e., the training set). We use this trained classifier to predict on the remaining $(100-y)$ percent of the data (i.e., the testing set), measuring the accuracy and area under receiver operating characteristic curve (AUROC). }

\textcolor{black}{In summary, for a given $y$, for each balanced dataset, we have 10 accuracy and 10 AUROC scores, corresponding to the 10 data splits; for each measure, we compute the average over the 10 splits, obtaining a single accuracy and single AUROC. Then, for a given $y$, given the single accuracy and single AUROC for each balanced dataset, i.e., given 10 accuracy and 10 AUROC scores for the 10 balanced datasets, for each measure, we compute the average over the 10 balanced datasets to obtain a final accuracy and a final AUROC score for that $y$. In our evaluation, we vary $y$ from 10 to 90 in increments of 10; each variation is called a ``$y$ percent training test''. This allows us to test how the amount of training data affects the results, which is important because in many real-world applications, only a small amount of data may be available for training.}

\textcolor{black}{Only if the average accuracy and AUROC are high, i.e., if TARA-TS accurately predicts functionally (un)related nodes to be functionally (un)related, does it make sense to use TARA-TS to create an alignment for protein functional prediction. If this is the case, we create an alignment as follows. Given one $y/(100-y)$ split and the classifier trained on it, we take every node pair from the testing set that is predicted as functionally related and add it to the alignment. Here, it is important to only use the testing set for the alignment. This way, there is no overlap between node pairs in the testing set and node pairs in the training set. Consequently, this avoids any circular argument. For simplicity, we do not repeat this process for all data splits, as we found that the split choice had no major effect on the classification performance. We only use the ``first'' one, which in our implementation corresponds to a starting seed of 0 for Python's random number generator when performing sampling. We have a total of 270 alignments, corresponding to all combinations of the 3 TARA-TS versions, the 9 percent training tests, and the 10 balanced datasets.}

\subsection{Using an alignment for protein functional prediction} \label{sec:methods-pfp}


An ultimate goal of biological NA is across-species protein functional prediction, so each NA method must be evaluated in this context. For a given ground truth-rarity dataset (atleast$k$-EXP at the $r$ rarity threshold), we consider GO terms that annotate at least two yeast proteins and at least two human proteins; these minimums are required to be able to make predictions for the GO term. Then, we use a (TARA-TS's or an existing method's) alignment in an established protein functional prediction framework  \citep{meng2016local}, as follows. In the alignment, for each protein $u$ that is annotated by at least $k$ GO term(s), such that the GO term(s) annotate $r$ or fewer proteins (i.e., for each protein for which a prediction can actually be made), $u$'s true GO term(s) are hidden. Then, for each GO term $g$, the framework determines if the alignment is significantly ``enriched'' in $g$. The hypergeometric test is used for this, calculating if the number of aligned node pairs in which the aligned proteins share $g$ is significantly high ($p$-value less than 0.05 \citep{meng2016local}). If so,  node $u$ is predicted to be annotated by GO term $g$. Repeating for all applicable proteins and GO terms results in the final list of predicted protein-GO term associations. From this prediction list, the framework calculates the precision (percentage of the predictions that are in a given ground truth-rarity dataset) and recall (percentage of the protein-GO term association from a given ground truth-rarity dataset that are among the predictions).


\section{Results} \label{sec:results}
\subsection{\textcolor{black}{Comparison of TARA-TS versions}}\label{sec:results-tarats-vers}

\noindent \textbf{Classification.} Here, we study classification performance of the three TARA-TS versions \textcolor{black}{(graphlets, node2vec, metapath2vec)} and TARA, i.e., how correctly they predict as functionally (un)related the protein pairs from testing data in a given $y$ percent training test. We would ideally do this on all nine ground truth-rarity datasets. However, two of them, atleast3-EXP at the 50 and 25 thresholds, are too small for TARA-TS and TARA to perform any classification on; inability to learn on small datasets is a drawback of machine learning methods in general, not just TARA-TS and TARA. Thus, we have seven viable ground truth-rarity datasets.

\textcolor{black}{Due to space constraints, we discuss the effect of various parameters ($k$ in atleast$k$-EXP, GO term rarity threshold, and $y$ percent training test) on the classification performance of a given TARA-TS version, for each version, in Supplementary Section S1.1. Instead, here we focus on comparing the three TARA-TS versions and TARA. 
}

\begin{figure*}[ht]
    \captionsetup{position=top}
    \advance\leftskip-2cm
    \advance\rightskip-2cm
      \subfloat[\textcolor{white}{@@@@@@@@@@@@@@@@}]{\includegraphics[width=0.29\textwidth, valign=t]{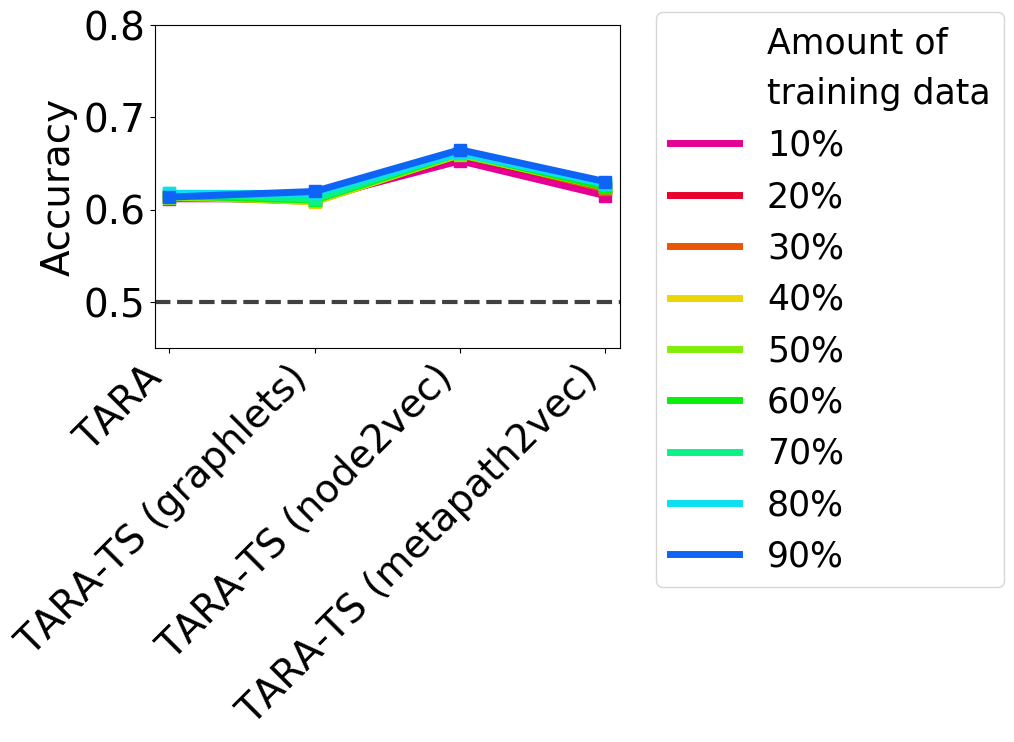}}\hspace{0.0cm}%
    \hspace{-0.1cm}     
      \subfloat[\textcolor{white}{@@@@@@@@@}]{\includegraphics[width=0.18\textwidth, valign=t]{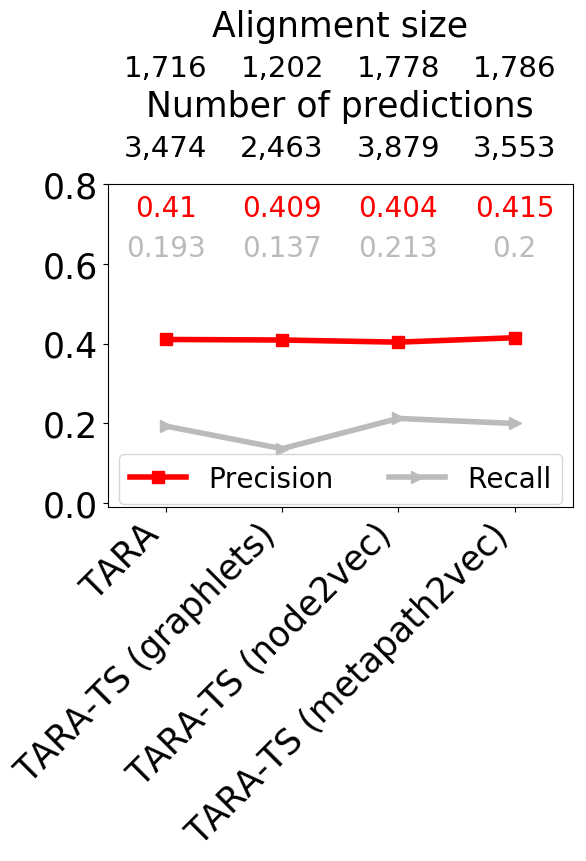}}\hspace{0.0cm}%
      \hspace{-0.13cm}     
        \subfloat[\textcolor{white}{@@@@@@@@@@@@@@}]{\includegraphics[width=0.27\textwidth, valign=t]{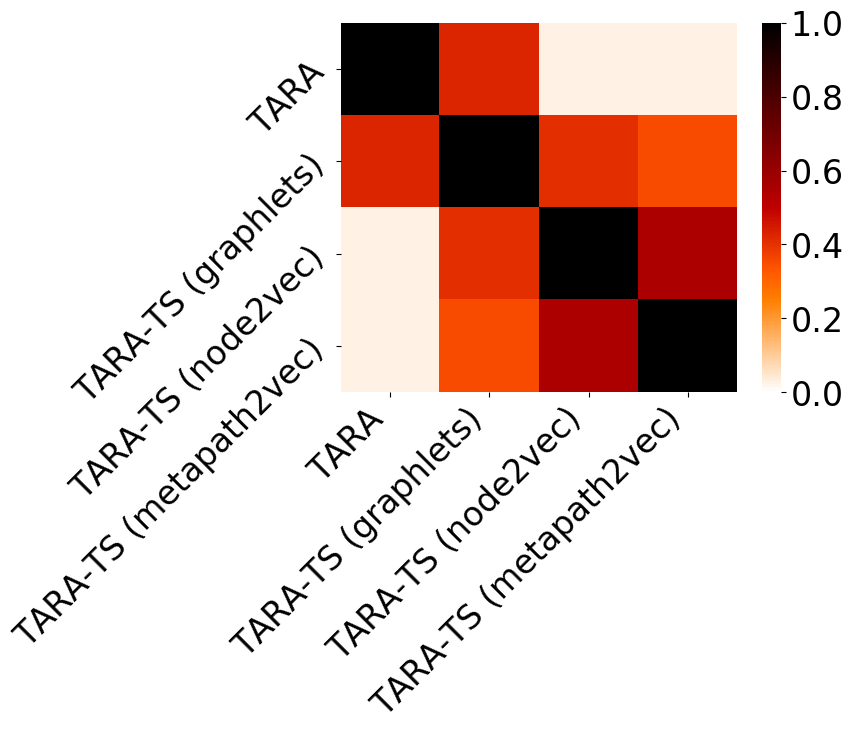}}\hspace{0.0cm}%
        \hspace{-0.13cm}     
        \subfloat[\textcolor{white}{@@@@@@@@@@@@@@}]{\includegraphics[width=0.27\textwidth, valign=t]{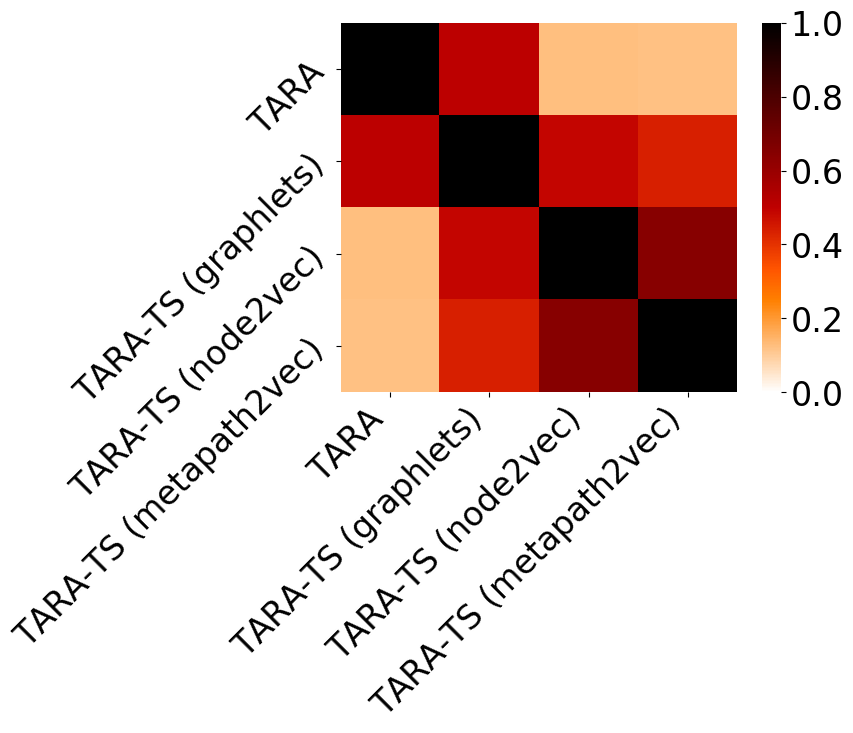}}%
\caption{\label{fig:tarats-comparisons} \textcolor{black}{Comparison of the three TARA-TS versions and TARA, for GO term rarity threshold 25 and ground truth dataset atleast1-EXP, in terms of: \textbf{(a)} classification accuracy, \textbf{(b)} protein functional prediction accuracy, \textbf{(c)}  overlap between  aligned yeast-human protein pairs, and \textbf{(d)}  overlap between predicted protein-GO term associations. In panel (b), the alignment for e.g., TARA contains 1,716 aligned protein pairs and predicts 3,474 protein-GO term associations. In panels (c)-(d), the pairwise overlaps are measured via the Jaccard index. Panel (a) encompasses all $y$ percent training tests. Panels (b)-(d) are for the 90\% training test.  Results for the other ground-truth rarity datasets and percent training \textcolor{black}{tests} are shown in Supplementary Figs. S1-S7.}}
\end{figure*}

\textcolor{black}{We expect all TARA-TS versions to have higher  accuracy and AUROC than TARA, as they extract topology plus sequence features from the integrated yeast-human network, unlike TARA, which extracts topology features only within each individual network. However, we find that this is not always the case (Fig. \ref{fig:tarats-comparisons}(a) and Supplementary Figs. S1-S2): 
(i) The \emph{relative} accuracy change of TARA-TS (graphlets) over TARA ranges from -3\% (decrease) to 5\% (increase), depending on the atleast$k$-EXP ground truth dataset, GO term rarity threshold, and $y$ percent training test, with an average change of 0\%; and its relative AUROC change ranges from -3\% to 5\%, with an average change of 1\%. 
(ii) TARA-TS (node2vec) does always improve over TARA though. Its relative accuracy change over TARA  ranges from 6\% to 27\%, with an average change of 14\%; and its relative AUROC change ranges from 9\% to 32\% with an average change of 16\%. 
(iii) As for TARA-TS (metapath2vec), we also see improvement over TARA, though not as large as for TARA-TS (node2vec). In particular, the relative accuracy change of TARA-TS (metapath2vec) over TARA ranges from -1\% to 14\% with an average change of 6\%; and its relative AUROC change ranges from 2\% to 15\%, with an average change of 7\%. }

Overall, we find that in terms of classification performance, TARA-TS (node2vec) to perform the best, followed by TARA-TS (metapath2vec), and followed by TARA-TS (graphlets) and TARA that are tied; all four perform significantly better than at random (Supplementary Figs. S1-S2).

\noindent \textbf{Protein functional prediction.} \textcolor{black}{Here, we evaluate the protein functional prediction accuracy of alignments of the three TARA-TS versions and TARA. For simplicity, we consider a subset of all nine $y$ percent training tests, focusing on the extremes (10 and 90) and the middle (50), because classification accuracy does not vary much between the different percent training options. Recall that classification cannot be performed on two (small) ground truth-rarity datasets, atleast3-EXP at thresholds 50 and 25, so no alignments exist for them, and thus protein functional prediction is not possible. So, for each TARA-TS version and TARA, we have 21 evaluation tests, resulting from combinations of the three selected $y$ values and the seven viable (as opposed to all nine) ground truth-rarity datasets.}

\textcolor{black}{First, we study the performance of the three TARA-TS versions. For each version, we expect that as we increase the amount of training data (10 to 50 to 90), precision will increase and recall will decrease. This is because a larger training dataset likely means that the classifier will be better (increasing precision), but will lead to a smaller testing dataset and thus smaller alignments and fewer predictions (decreasing recall). We expect that as $k$ increases (in our atleast$k$-EXP ground truth datasets), precision will increase and recall will decrease. This is because at higher $k$, TARA-TS will be trained on higher-quality data (increasing precision), but there will be less data overall, resulting in smaller alignments and fewer predictions (decreasing recall). We expect that as we consider rarer GO terms, precision will increase and recall will decrease. Rarer GO terms may be more meaningful \citep{hayes2017sana, gu2019data}, so the data will be of higher quality (increasing precision), but there will be less of it overall (decreasing recall). Indeed, we observe all of these expected trends for all TARA-TS versions (Supplementary Fig. S3-S5).}

\textcolor{black}{Second, we compare the performance of the three TARA-TS versions and TARA. Shockingly, all four methods yield almost equal protein functional prediction accuracy (Fig. \ref{fig:tarats-comparisons}(b) and Supplementary Figs. S3-S5). This is unexpected, as TARA-TS uses sequence information that TARA does not. We take a closer look at the alignments and predictions made by each method to see if the different methods are aligning the same nodes, or predicting the same protein-GO term associations. So, we investigate how much their alignments overlap (Fig. \ref{fig:tarats-comparisons}(c)), and how much their predictions overlap (Fig. \ref{fig:tarats-comparisons}(d)).  We find that the different methods are all aligning and predicting at least somewhat different information from each other. Yet, their predictions are equally accurate. Also, we find that TARA is more similar  to (i.e., overlaps the most with) TARA-TS (graphlets) than to TARA-TS (node2vec) and TARA-TS (metapath2vec), which makes sense since the former also uses graphlets to extract feature vectors like TARA, and the latter two do not. Also, TARA-TS (node2vec) and (metapath2vec) are more similar to each other than to the other methods, which makes sense since they use a similar random walk-based feature extraction method.}


\textcolor{black}{It is surprising that TARA-TS (graphlets) does not improve upon TARA, i.e., that the additional sequence information does not improve upon only topological information. A reason may be that the 
across-network sequence information complements, rather than enhances, the within-network topology information. Some of the predictions made by TARA-TS (graphlets), specifically those that overlap with TARA's, may be due to the within-network topology information used by both methods, and the remaining predictions made by TARA-TS (graphlets) may be due to the across-network sequence information, which is not used by TARA.}

\textcolor{black}{Also, it is surprising that TARA-TS (metapath2vec) does not improve upon TARA-TS (node2vec). Both use a similar random walk-based embedding process, but metapath2vec additionally accounts for the heterogeneous information in the integrated network. The lack of improvement may be because the additional information captured by the considered metapaths is not useful in this task, or because constraining random walks by node type leads to less neighborhood structure being explored. For example, at some point in a random walk, a human node may have many human neighbors, but the walk is forced to move to a yeast node due to the metapath constraints. Then, the neighborhood of that human node will not be well explored. However, because the number of possible metapaths to test in order to find the best one(s) is exponential with respect to the length of the path, it is not feasible to test every possibility, even for short lengths. Thus, an efficient way of selecting appropriate metapaths for a given network would be necessary to continue to pursue metapath-based embedding methods for this task. However, to our knowledge no such selection process exists, which is why we do not pursue this problem beyond the metapaths we have considered.}

\textcolor{black}{Because TARA-TS (node2vec) not only yields the best classification performance, i.e., predicts functional (un)relatedness the best out of all TARA-TS versions but also  captures the most novel protein functional information compared to TARA (i.e., the predictions it makes overlap the least to those made by TARA out of all TARA-TS versions), we continue only with TARA-TS (node2vec) as the selected TARA-TS version.}


\subsection{TARA-TS versus TARA in the task of protein functional prediction: toward TARA++} \label{sec:tara-ts-combination}

\begin{figure}[ht!]
 \centering    
(a)\includegraphics[width=0.25\textwidth,valign=t]{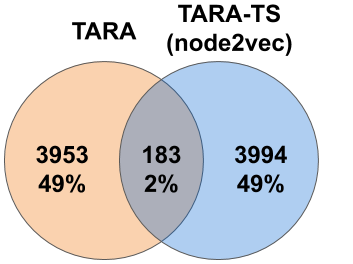} 
(b) \hspace{0.1cm} \includegraphics[width=0.3\textwidth,valign=t]{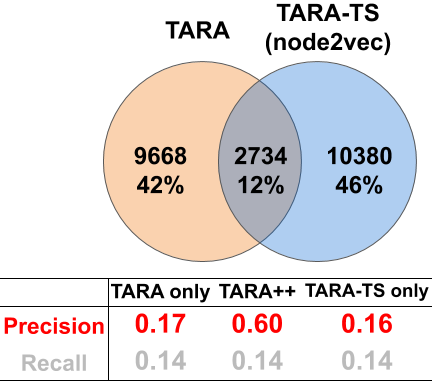}
 \caption{\label{fig:tara-vs-tarats} Comparison of the selected TARA-TS version and TARA, for GO term rarity threshold 50, ground truth dataset atleast1-EXP, and the 90\% training test, in terms of overlap between their: \textbf{(a)} aligned yeast-human protein pairs and \textbf{(b)} predicted protein-GO term associations. In panel (b),  precision and recall are shown for each of the three prediction sets captured by the Venn diagram; TARA++'s predictions are those in the overlap. The overlaps are for one of the 10 balanced datasets; so, the alignment size and prediction number of a method may differ from those in Fig. \ref{fig:tarats-comparisons}(b), where the statistics are averaged over all balanced datasets.  Results for the other ground truth-rarity  datasets are shown in Supplementary Figs. S8-S9.}
\end{figure}

\textcolor{black}{Focusing on TARA-TS (node2vec) as the selected TARA-TS version (i.e., simply as TARA-TS), we zoom into the comparison between it and TARA. The two methods have different alignments and make different predictions (Fig. \ref{fig:tara-vs-tarats}), so how can they still have similar protein functional prediction accuracy?} To answer this, we look at the precision and recall of predictions made by both methods, only those predictions made by TARA-TS but not TARA, and only those predictions made by TARA but not TARA-TS (Fig. \ref{fig:tara-vs-tarats}(b) and Supplementary Fig. S9). 
From this, we highlight two findings. First, graphlets, which  use only topological information,  perform as well as network embedding features that use both  topological and sequence information. This is supported by the fact that predictions made only by TARA and only by TARA-TS produce similar accuracy in almost all evaluation tests. This motivates the need to develop better graphlet-based methods for integrated networks as future work. Second, predictions made by both methods are significantly more accurate than predictions made by any one method alone. In a sense, their overlap is integrating state-of-the-art research across the computational biology and social network domains, by combining TARA's graphlet-based topology-only features with TARA-TS's embedding-based topology-and-sequence features. So, the overlapping predictions combine the strengths of both domains.

Because the overlap of TARA-TS and TARA has such high prediction accuracy, we take it as our new TARA++ method, which we consider further. Then, to simplify comparisons between TARA++ and existing NA methods, we choose a representative percent training test (either TARA++10, TARA++50, or TARA++90) for each of the seven viable ground truth-rarity datasets. In other words, we go from 21 ``TARA++ versus existing methods'' evaluation tests to seven. Generally, for each viable ground truth-rarity dataset, we try to choose the percent training test that has both high precision (predictions are accurate) and a large number of predictions (we uncover as much of biological knowledge as possible), as these represent TARA++'s best results. So, we choose TARA++90 for all ground truth-rarity datasets except atleast2-EXP at the 50 and 25 rarity thresholds, where we choose TARA++10. Henceforth, we refer to all of the selected TARA++ versions simply as TARA++.


\subsection{TARA++ versus existing NA methods in the task of protein functional prediction}


We compare  TARA++'s predictions  against those of two most fairly comparable state-of-the-art methods, TARA and PrimAlign. Also, we consider predictions resulting from using only sequence information, Sequence. Here, we treat the 55,594 anchor links by themselves  as the alignment; as no topological information is used,  this is not an NA method. With TARA and Sequence, we can separately analyze each aspect, i.e., within-network topological information only and across-network sequence information only, and evaluate how each compares to our integrative TARA++. (TARA++'s predictions are by definition a subset of TARA's predictions, and so we expect TARA++ to have higher precision but lower recall than TARA.) With PrimAlign, we can evaluate how this  integrative but unsupervised method compares to our also integrative but supervised TARA++. 
Importantly,  TARA and PrimAlign were already shown to outperform many previous NA methods (Section \ref{sec:rel-work}). So, comparing to these two methods is sufficient. \textcolor{black}{Also, keep in mind that like with TARA, a theoretical precision of 1 is not practically possible with TARA++. This is because for a given training/testing split, TARA++ uses a part (up to 90\%) of the ground truth functional data for training, and so for that split, it is impossible to make predictions for the training data portion.}


We believe that precision is more important than recall. This is because for potential wet lab validation of predictions, it is more important to have fewer but mostly correct predictions (e.g., 90 correct out of 100 made) than a greater number of mostly incorrect predictions (e.g., 300 correct out of 1000 made). While in the latter example more predictions are correct, leading to higher recall, many more are also incorrect, leading to lower precision. We do not entirely discount recall though, as it  still brings  value. 

\begin{figure*}[ht!]
     \centering
    \advance\leftskip-2cm
    \advance\rightskip-2cm
      \sidesubfloat[]{\includegraphics[width=0.25\textwidth, valign=t]{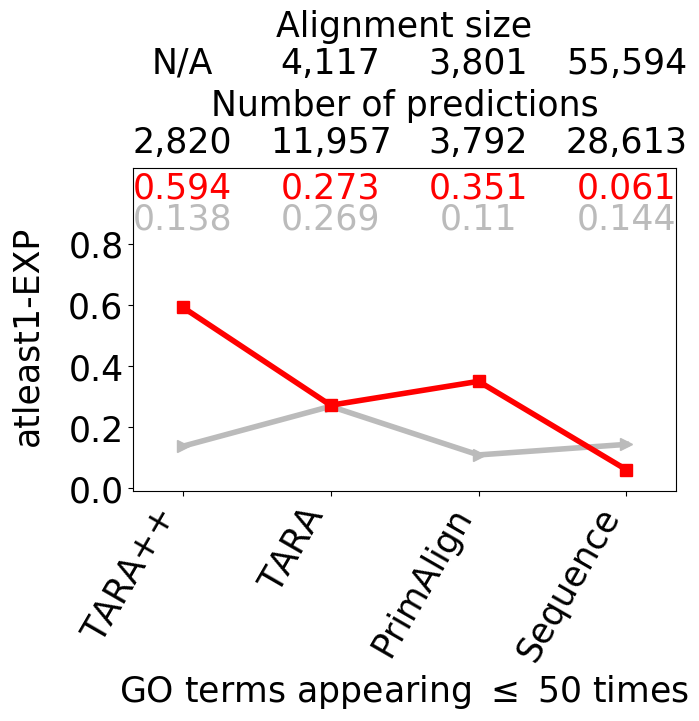}\label{fig:tarapp-b}}\hspace{1.2cm}%
      \sidesubfloat[]{\includegraphics[width=0.25\textwidth, valign=t]{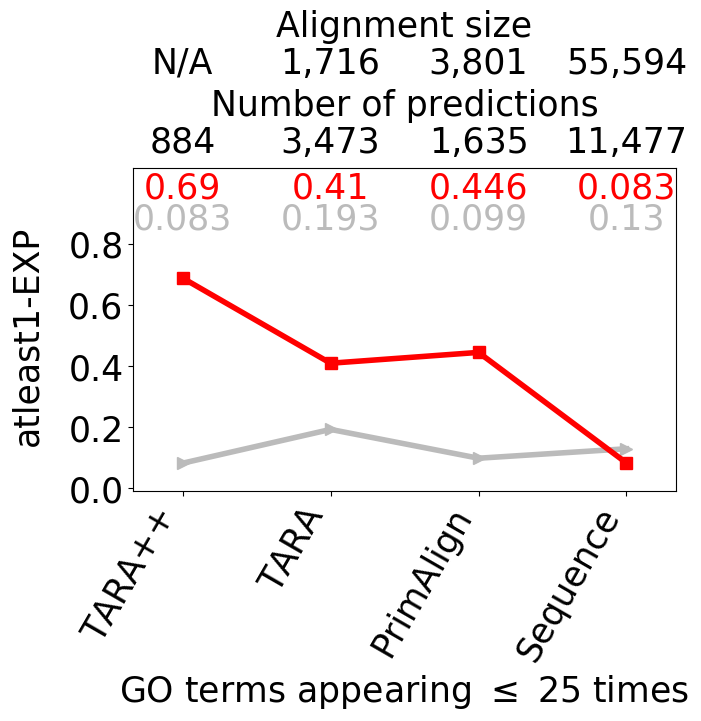}\label{fig:tarapp-c}}\hspace{1.2cm}%
      \sidesubfloat[]{\includegraphics[width=0.35\textwidth, valign=t]{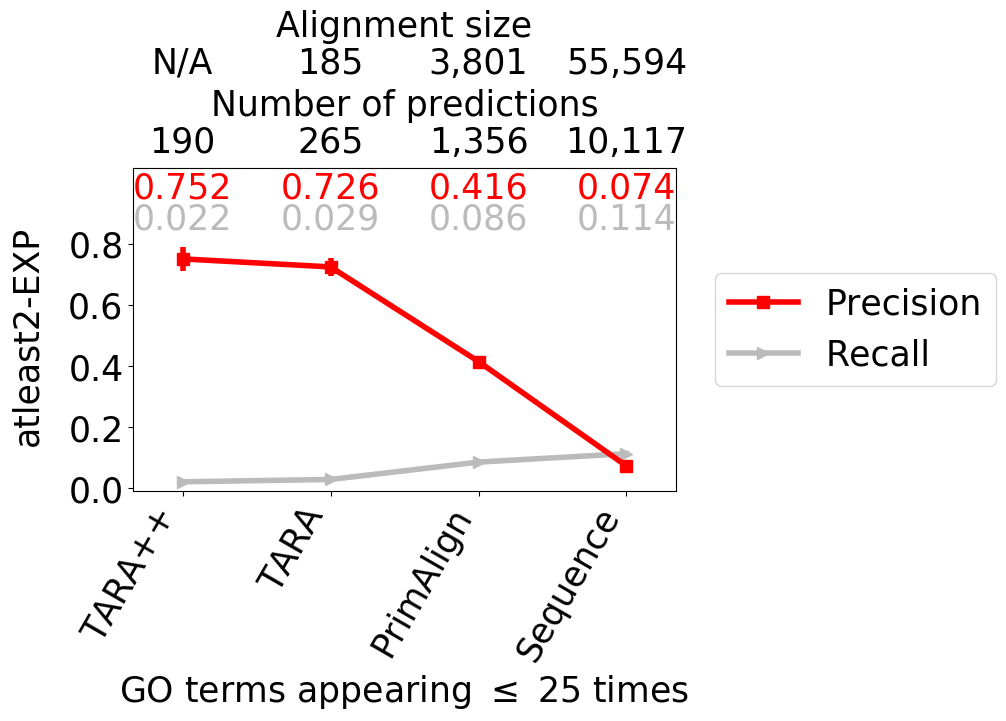}\label{fig:tarapp-f}}%

    
    
 \caption{\label{fig:tarapp-vs-all} Comparison of TARA++ and three existing methods  in the task of protein functional prediction, for rarity thresholds \textbf{(a)} 50 and \textbf{(b, c)} 25, and for ground truth datasets \textbf{(a, b)} atleast1-EXP and \textbf{(c)} atleast2-EXP. The alignment size (the number of aligned yeast-protein pairs) and number of functional predictions (predicted protein-GO term associations) are shown for each method, except \textcolor{black}{that TARA++ does not have an alignment \emph{per se}. i.e., TARA++ comes from the overlap of \emph{predictions} made by TARA and TARA-TS; hence the ``N/A''s.}
 \textcolor{black}{For TARA++ and TARA, results are averages over all balanced datasets; the standard deviations are small and thus invisible.}   Results for the other ground truth-rarity datasets are shown in Supplementary Fig. \textcolor{black}{S10}.}
\end{figure*}

Our key results are as follows (Fig. \ref{fig:tarapp-vs-all} and Supplementary Fig. S10). In terms of precision, TARA++ is the best for 6/7 ground truth-rarity datasets. It is only slightly inferior to PrimAlign for 1/7 datasets (atleast1-EXP for ALL GO terms), but TARA++ has much higher recall than PrimAlign on this dataset. Speaking of recall, TARA is  expected to always outperform TARA++, and this is what we observe. Of the remaining existing methods, TARA++ is the best for 2/7 datasets -- atleast1-EXP at the ALL and 50 rarity thresholds -- even though TARA++ makes much \emph{fewer} predictions than the next best method, Sequence. For the other datasets, TARA++'s recall is lower than that of PrimAlign and Sequence. This is expected, since TARA++ makes fewer predictions than the other methods. Importantly, the difference in recall between TARA++ and every other method is relatively small, for example only 0.06 lower on average compared to TARA, while TARA++ is much better in terms of precision than every other method, for example 0.2 greater on average compared to TARA. As discussed above, such a trade-off between precision and recall is worth it for our task.

We see that the precision of TARA++ is much greater than simply the sum of precision from TARA and Sequence, suggesting that integrating within-network topological and across-network sequence information has compounded effects. This further highlights the need for such approaches.


Finally, we look at the time it takes to obtain an alignment for TARA-TS, TARA, and PrimAlign, for the ALL GO term rarity threshold, which has the most data and  is thus the worst case time-wise out of all thresholds. As TARA++ comes from the intersection of TARA-TS's and TARA's results, its time is either the maximum or sum of TARA-TS's and TARA's, if the two are run at the same time or one after the other, respectively. Our findings are as follows (also, see Supplementary Table S1). 

For TARA-TS, as $k$ (in the atleast$k$-EXP) increases, the time to produce an alignment is expected to decrease, as there is less (but higher-quality) data overall, and thus less data to train on. This is what we observe. 
When comparing TARA-TS and TARA, the former  is faster, and this comes from the feature computation time, as both use the same supervised framework. TARA-TS's node2vec computation  is expectedly faster than TARA's graphlet counting even when using Orca  for two reasons. 
\textcolor{black}{First, the random walks produced by node2vec can be thought of as sampling the network structure, which is much faster than capturing the full network structure like graphlets do.}
\textcolor{black}{Second, node2vec is parallelized while Orca is not. Parallelization benefits node2vec a lot: the same number of random walks is performed for each node (parameter \texttt{-r:}), so no single node takes much longer than any other. However, for graphlet counting, nodes with e.g., high degrees are the limiting time factor, and so parallelization would not help as much.}
\textcolor{black}{Also note that TARA-TS's (and PrimAlign's) running time is missing the step of computing sequence-based anchor links; these anchors were precomputed and provided by the PrimAlign study. So, TARA-TS (and PrimAlign) has an unfair advantage over TARA.}
Despite this missing step, regardless of how TARA-TS and TARA are combined to form TARA++, PrimAlign will still be faster. However, it is about half as precise as TARA++. Even though TARA++ is slower, it is still practically feasible. So, the extra time is worth the almost doubling of precision.




\section{Conclusion}
TARA++ pushes the data-driven NA paradigm further. We showed that by integrating research knowledge across the computational biology and social network domains, TARA++ outperforms state-of-the-art NA methods in the task of protein functional prediction, an ultimate goal of NA. 
%
As TARA++ is the first data-driven NA method to integrate topological and sequence information, it is just a proof-of-concept approach. This work can be taken further. We found that graphlet-based features on the isolated networks (on topological information alone) perform as well as embedding-based features on the integrated network (on topological and sequence information combined), even though the latter (using more data) was expected to be better. So, developing a graphlet feature that would efficiently deal with an integrated network could  yield  further  improvements. This might include novel algorithms for 
speeding up counting of heterogeneous graphlets in large data. Heterogeneous graphlets, or heterogeneous network embedding features 
\textcolor{black}{other than metapath2vec},  
could \textcolor{black}{better} distinguish between different node/edge types in an integrated  network and thus only improve over the 
features considered in this study. Also, we focused on NA of static networks. However, research in NA of dynamic (e.g., aging- or disease progression-related) networks  is becoming popular \citep{vijayan2017alignment, vijayan2017aligning}. So, our framework can be adapted to such novel NA categories.

\section*{Funding}
National Science Foundation (NSF) CAREER CCF-1452795 award.

%
\bibliographystyle{natbib}
%
%

\end{document}


\title{Supplementary material: Data-driven biological network alignment that uses topological, sequence, and functional information}

\maketitle

\newcommand{\beginsupplement}{%
        \renewcommand{\figurename}{Supplementary Figure}
        \renewcommand{\tablename}{Supplementary Table}
        \renewcommand{\thetable}{S\arabic{table}}%
        \renewcommand{\thefigure}{S\arabic{figure}}%
        \renewcommand{\thesection}{S\arabic{section}}%
        
}

\newcount\suppl
\suppl0
\def\Ref#1{\ifnum\suppl<1 \ref{#1} \else S\ref{#1}\fi}

\beginsupplement

\section{Results}
\subsection{TARA-TS versus TARA in the classification context}

Here, we comment on the performance of TARA-TS; recall that we use TARA-TS to refer to any of TARA-TS (graphlets, node2vec, metapath2vec). For a fixed GO term rarity threshold, as $k$ in our atleast$k$-EXP ground truth datasets increases, we expect TARA-TS's (and TARA's) accuracy and AUROC to increase, as the condition for proteins to be functionally related becomes more stringent and thus the functional data becomes of higher quality. Also, for a fixed $k$, as we decrease the GO term rarity threshold (i.e., consider rarer GO terms), we expect accuracy and AUROC to increase, since rarer GO terms may be meaningful (Hayes and Mamano, 2017), again resulting in higher-quality data.  We find the former expectation to hold, for all GO term rarity thresholds (Supplementary Figs. S1-S2). However, for the latter expectation, we find that classification accuracy and AUROC somewhat decrease (Supplementary Figs. S1-S2). This may be because as rarer GO terms are considered, the amount of training data decreases, which is what could be causing performance decreases.

As we increase $y$, the amount of training data, we expect accuracy and AUROC to increase, as more data is used during classification. For accuracy, for TARA-TS (graphlets), we observe this for 6/7 ground truth-rarity datasets, although for 4/6 of the datasets, the increase is minimal ($\sim$1\%). In the remaining case, the accuracy increases until about 60\% training data, and then drops. For TARA-TS (node2vec), we observe this for 6/7 ground truth-rarity datasets, although for 4/6, the increase is minimal. In the remaining case, the accuracy increases until about 60\% training data, and then drops. For TARA-TS (metapath2vec), we observe this for all 7 ground-truth rarity datasets. 
For AUROC, for TARA-TS (graphlets), we observe the expected trend for all ground truth-rarity datasets, although for 4/7 of these datasets, the increase is minimal ($\sim$1\%). For both TARA-TS (node2vec) and TARA-TS (metapath2vec), we observe the expected trend for all ground-truth rarity datasets, although for 3/7 of these datasets, the increase is minimal. These unexpected trends (mostly minor increase of accuracy and AUROC even with large increase of $y$)  are promising though, because they mean that TARA-TS does not have to use a majority of the functional data for training to still obtain good results; even using only 10\% of the data seems to suffice.

\section{Supplementary figures and tables}

\begin{figure}[ht!]

    \begin{subfigure}[t]{0.29\textwidth}
        \includegraphics[width=\textwidth]{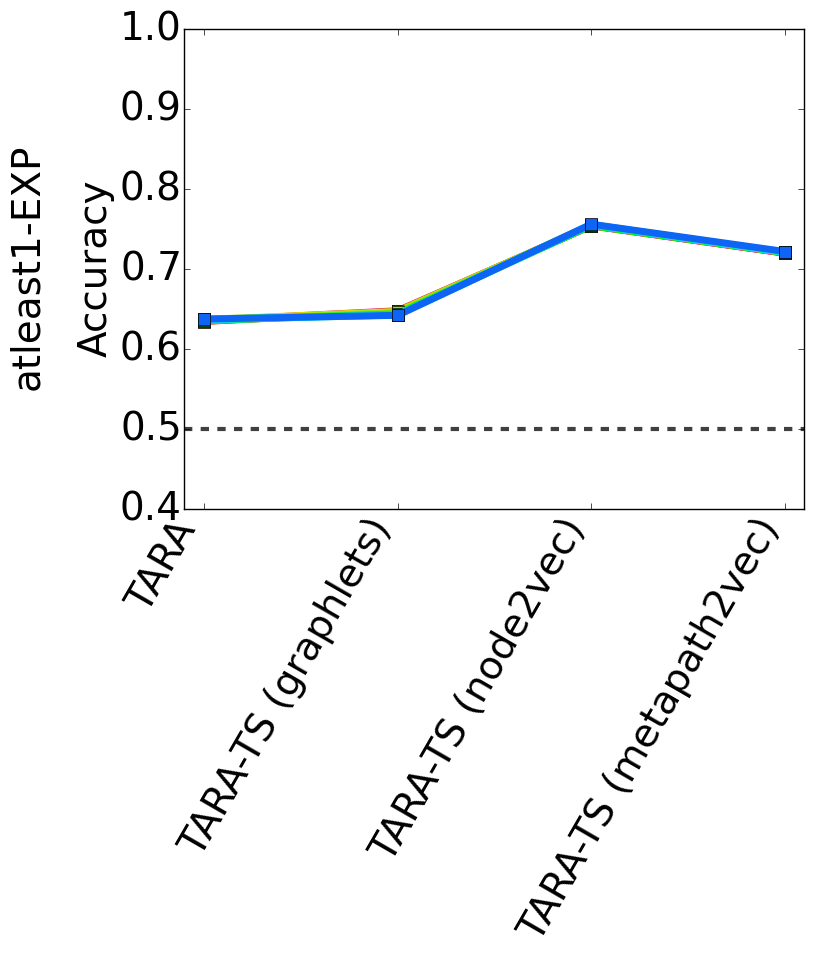}
        \subcaption{}
    \end{subfigure}
    \begin{subfigure}[t]{0.26\textwidth}
        \includegraphics[width=\textwidth]{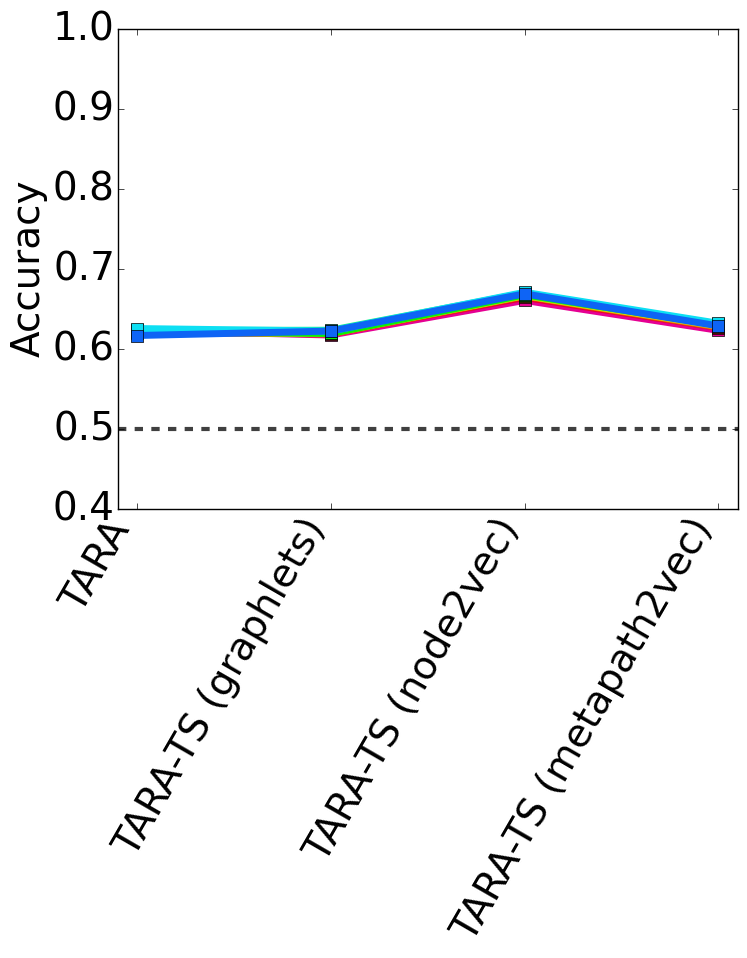}
        \subcaption{}
    \end{subfigure}
    \begin{subfigure}[t]{0.405\textwidth}
        \includegraphics[width=\textwidth]{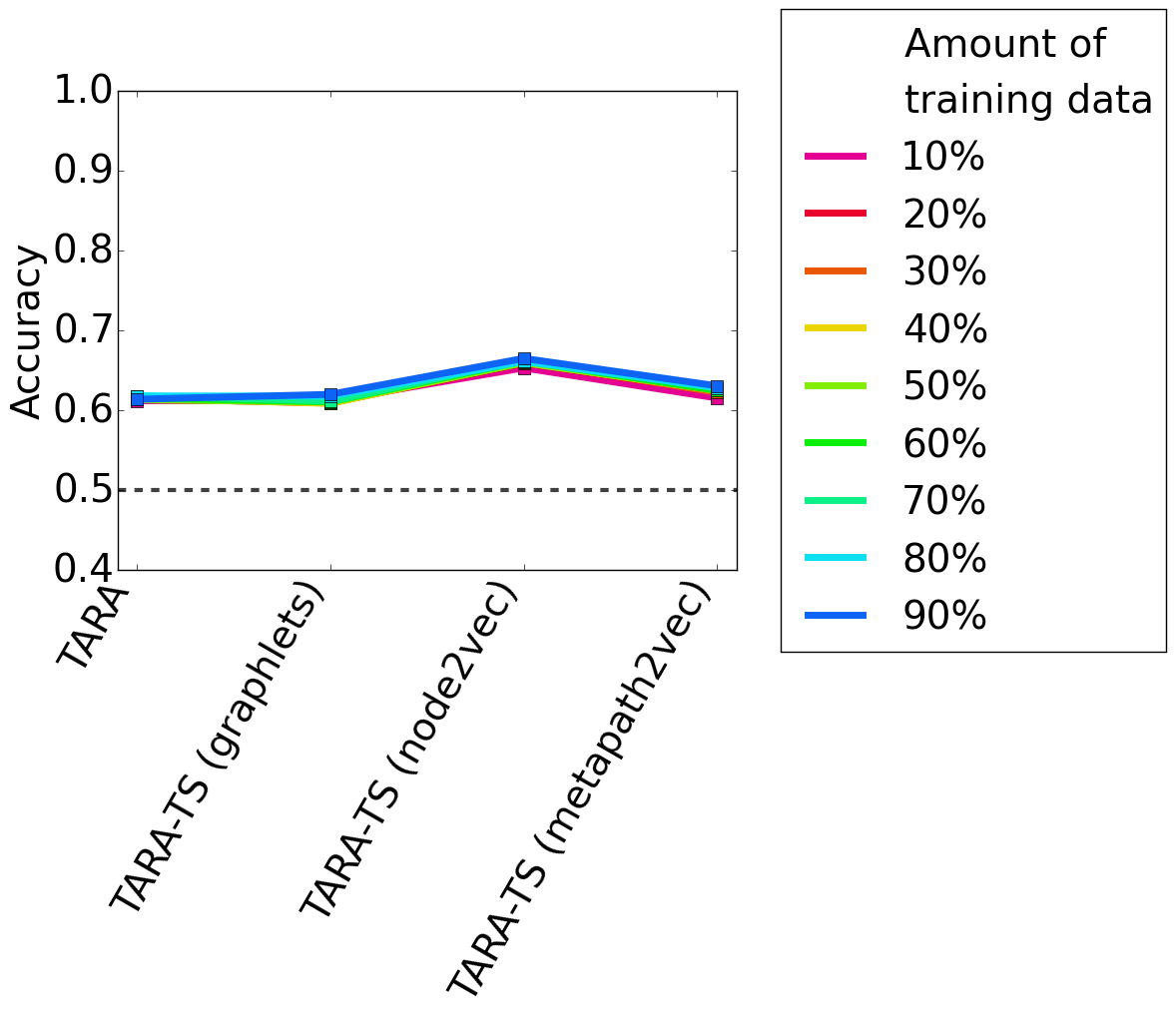}
        \subcaption{}
    \end{subfigure} \\
    \begin{subfigure}[t]{0.29\textwidth}
        \includegraphics[width=\textwidth]{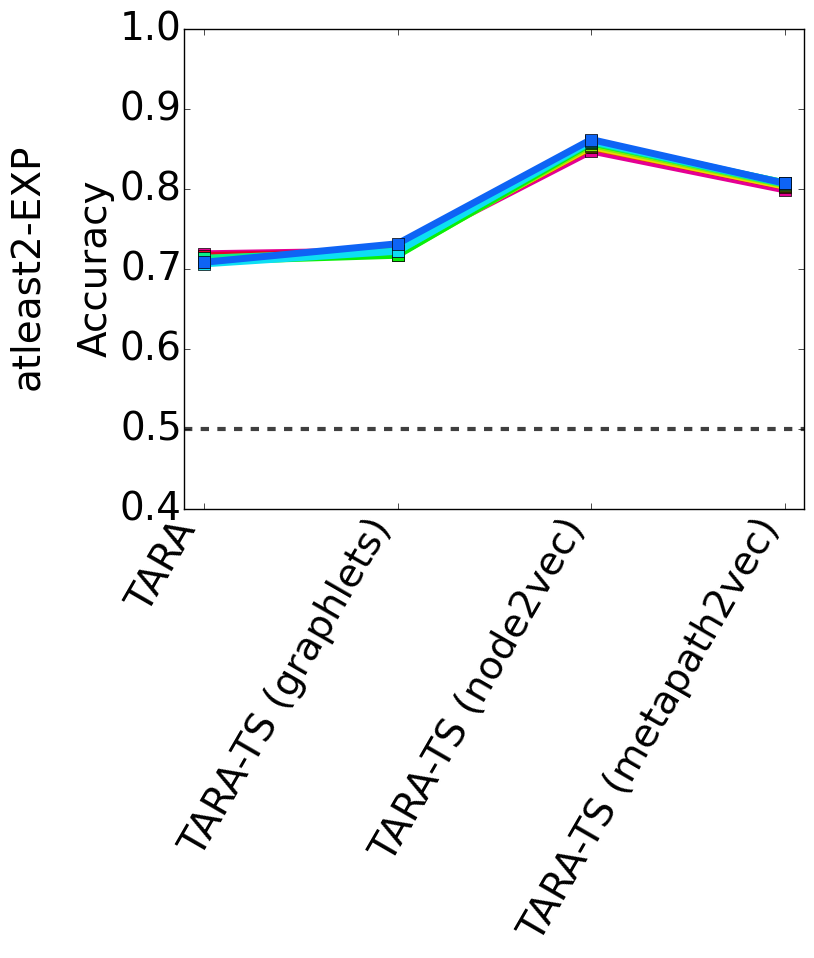}
        \subcaption{}
    \end{subfigure}
    \begin{subfigure}[t]{0.26\textwidth}
        \includegraphics[width=\textwidth]{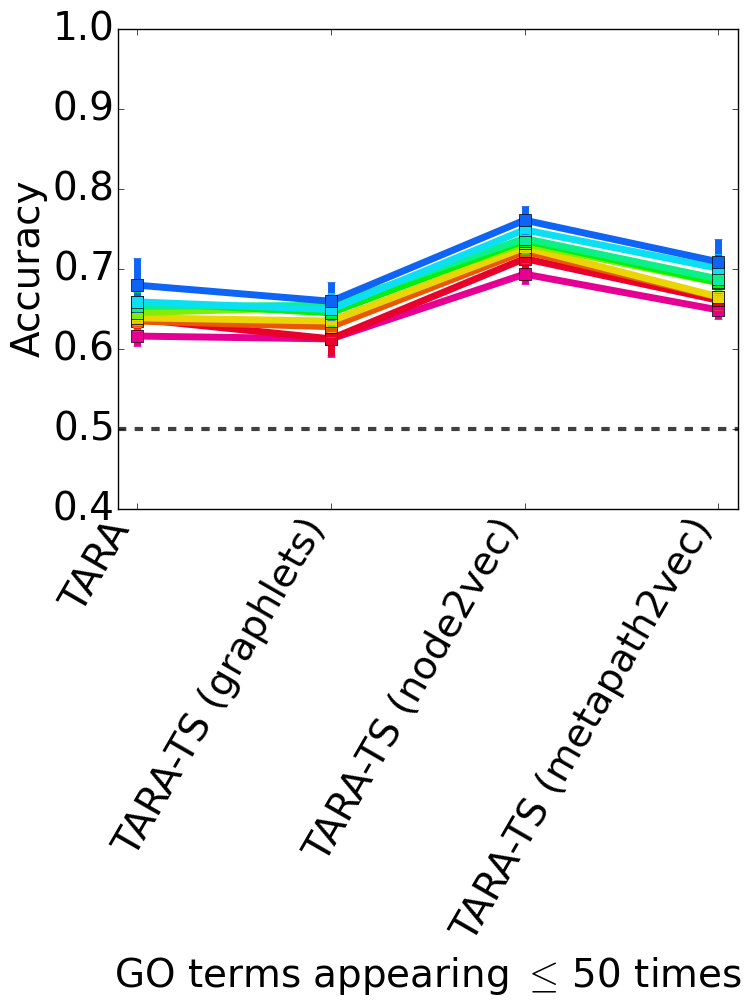}
        \subcaption{}
    \end{subfigure}
    \begin{subfigure}[t]{0.405\textwidth}
        \includegraphics[width=\textwidth]{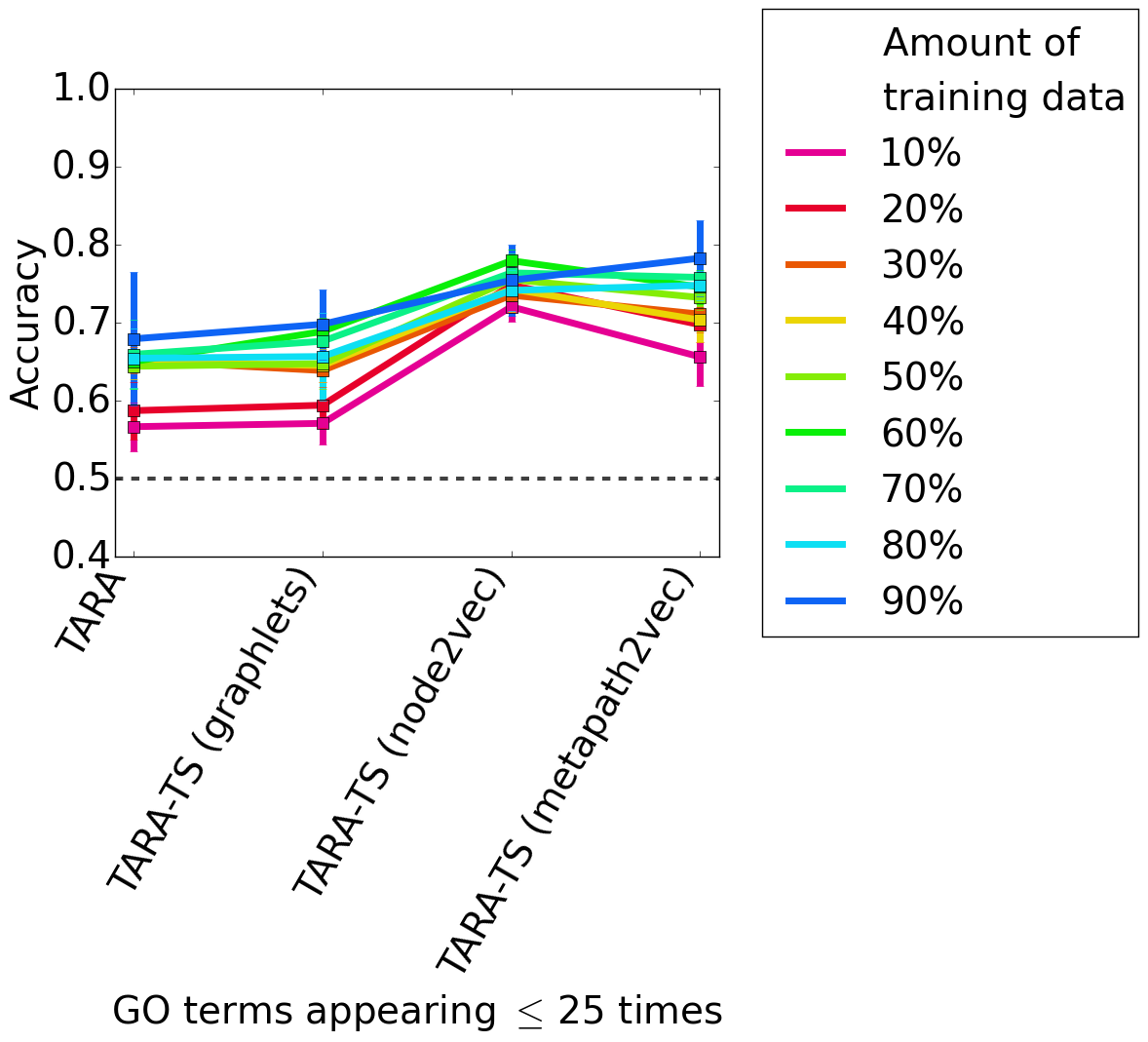}
        \subcaption{}
    \end{subfigure}

    \begin{subfigure}[t]{0.44\textwidth}
        \includegraphics[width=\textwidth]{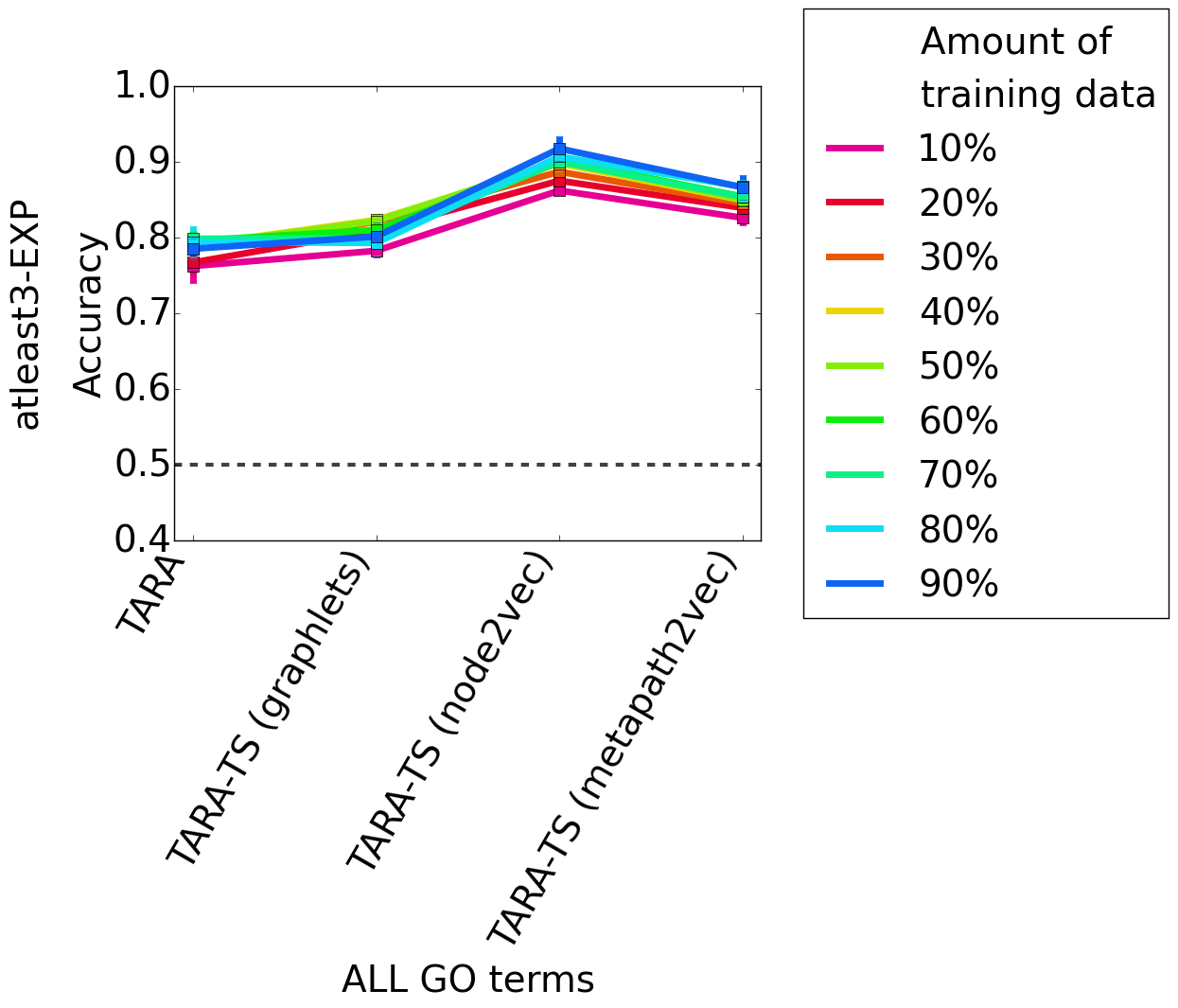}
        \subcaption{}
    \end{subfigure}

 \caption{\label{fig:supp-acc} \textcolor{black}{Average prediction accuracy of percent training tests for rarity thresholds \textbf{(a, d, g)} ALL, \textbf{(b, e)} 50, and \textbf{(c, f)} 25 using ground truth datasets \textbf{(a, b, c)} atleast1-EXP, \textbf{(d, e, f)} atleast2-EXP, and \textbf{(g)} atleast3-EXP. A dotted black line indicates the accuracy expected if the classifier makes random predictions. Qualitatively similar results for AUROC are shown in Supplementary Figs. S2.}}
\end{figure}

\begin{figure}[ht!]

    \begin{subfigure}[t]{0.29\textwidth}
        \includegraphics[width=\textwidth]{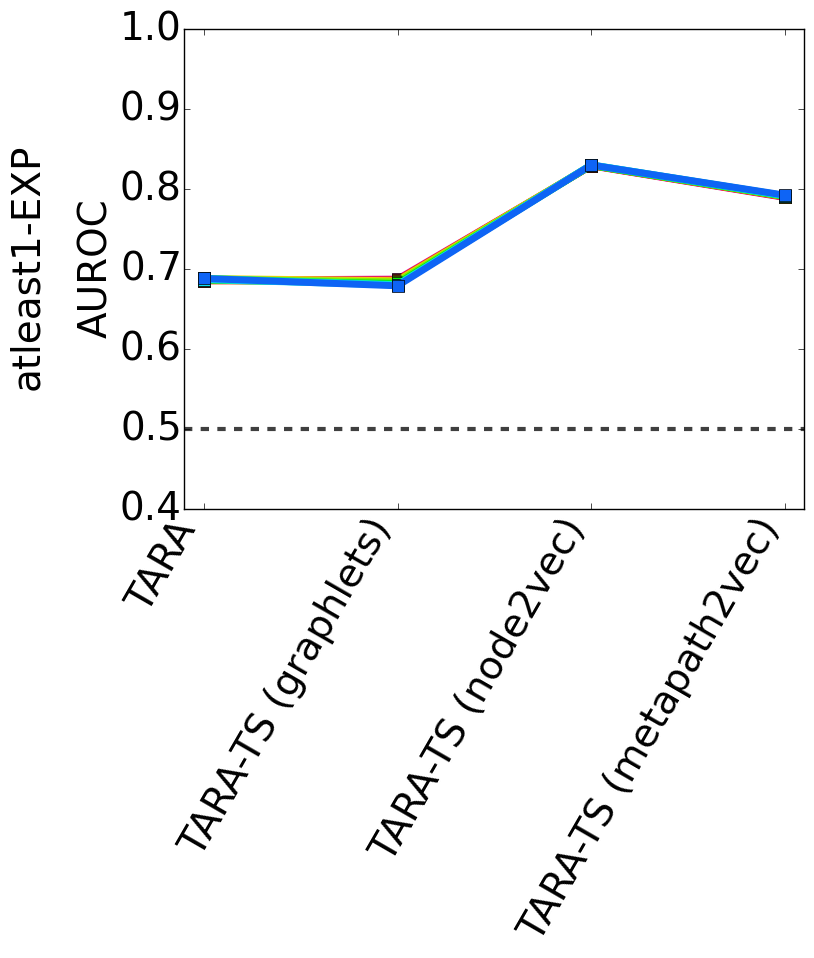}
        \subcaption{}
    \end{subfigure}
    \begin{subfigure}[t]{0.26\textwidth}
        \includegraphics[width=\textwidth]{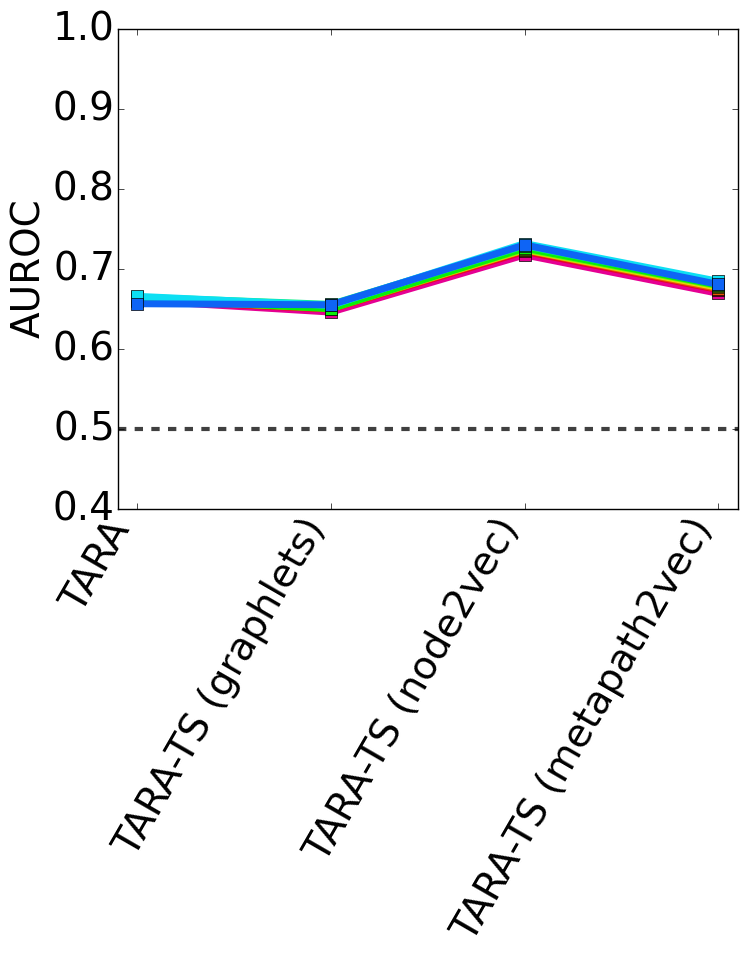}
        \subcaption{}
    \end{subfigure}
    \begin{subfigure}[t]{0.405\textwidth}
        \includegraphics[width=\textwidth]{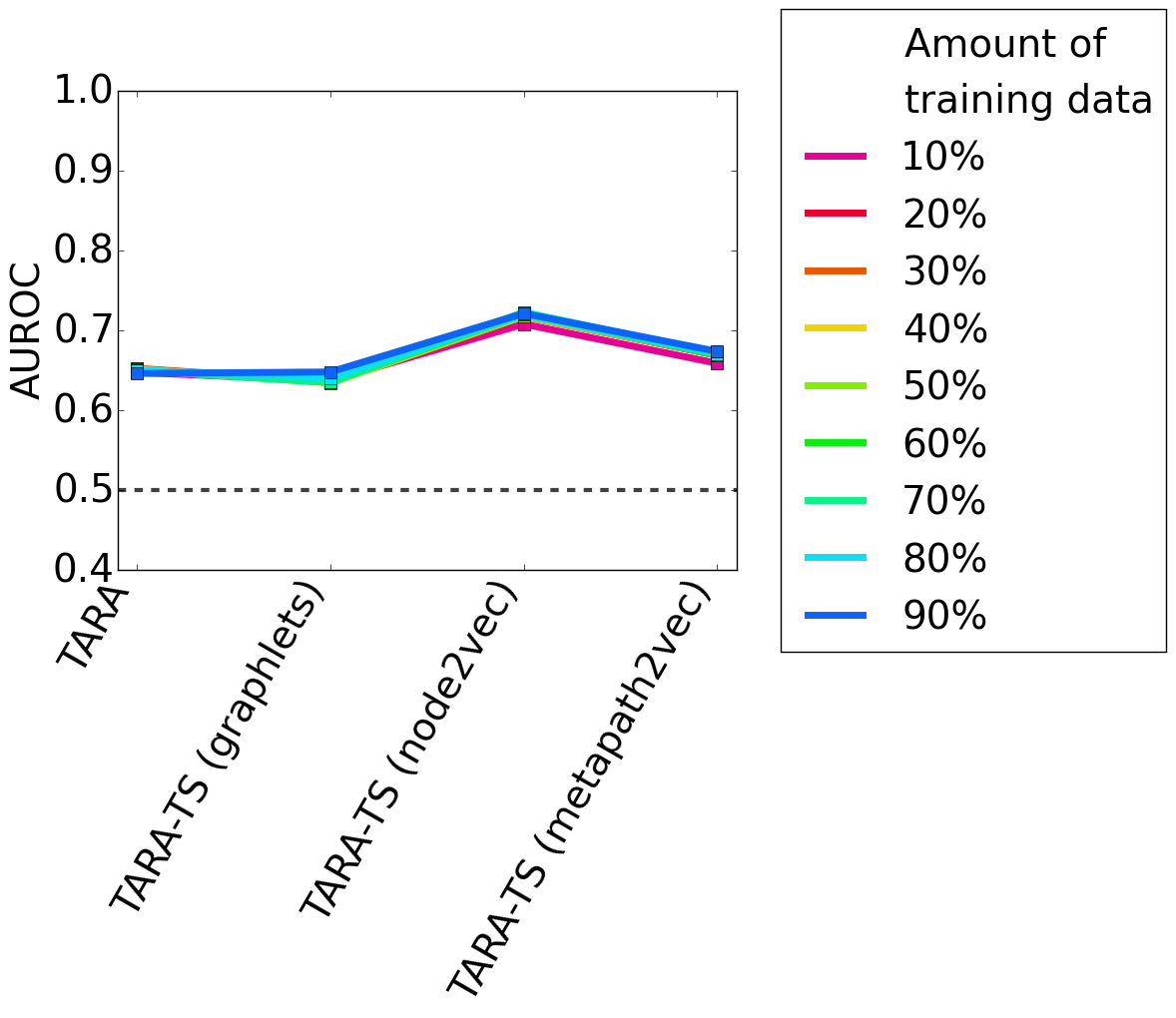}
        \subcaption{}
    \end{subfigure} \\
    \begin{subfigure}[t]{0.31\textwidth}
        \includegraphics[width=\textwidth]{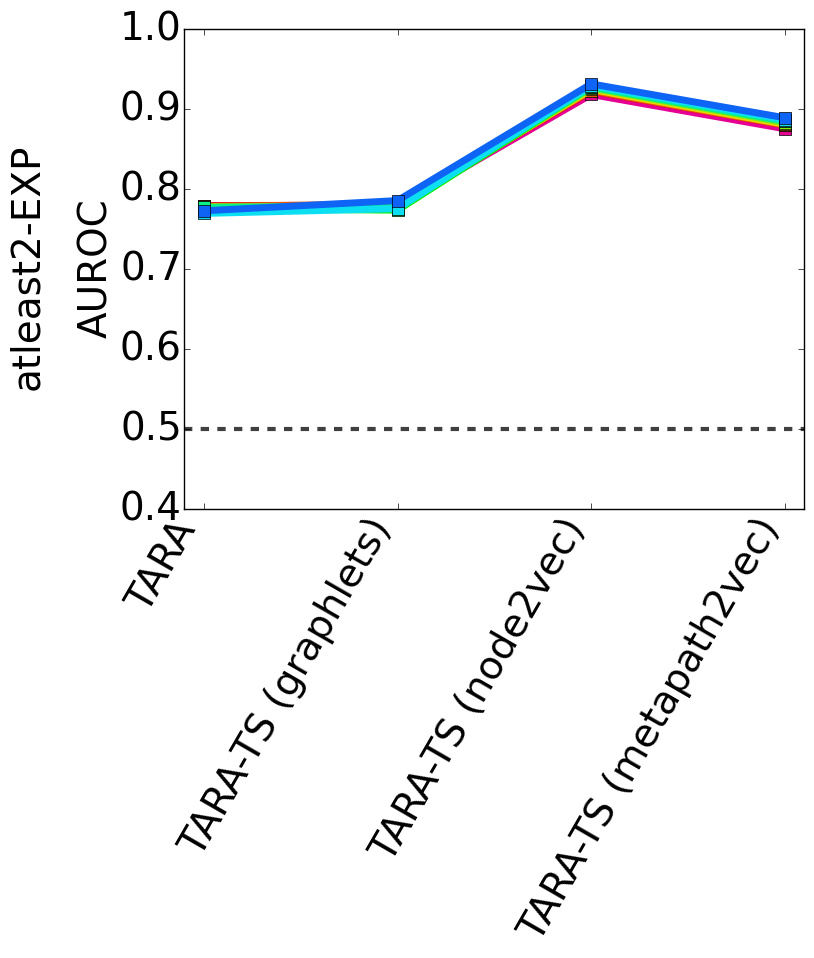}
        \subcaption{}
    \end{subfigure}
    \begin{subfigure}[t]{0.26\textwidth}
        \includegraphics[width=\textwidth]{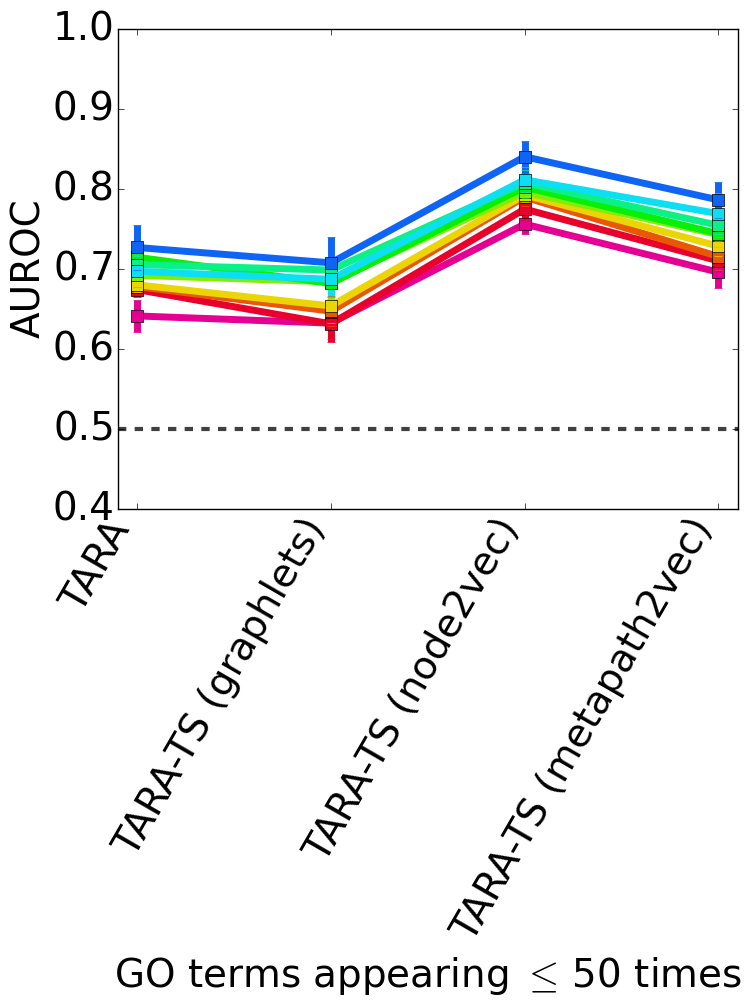}
        \subcaption{}
    \end{subfigure}
    \begin{subfigure}[t]{0.405\textwidth}
        \includegraphics[width=\textwidth]{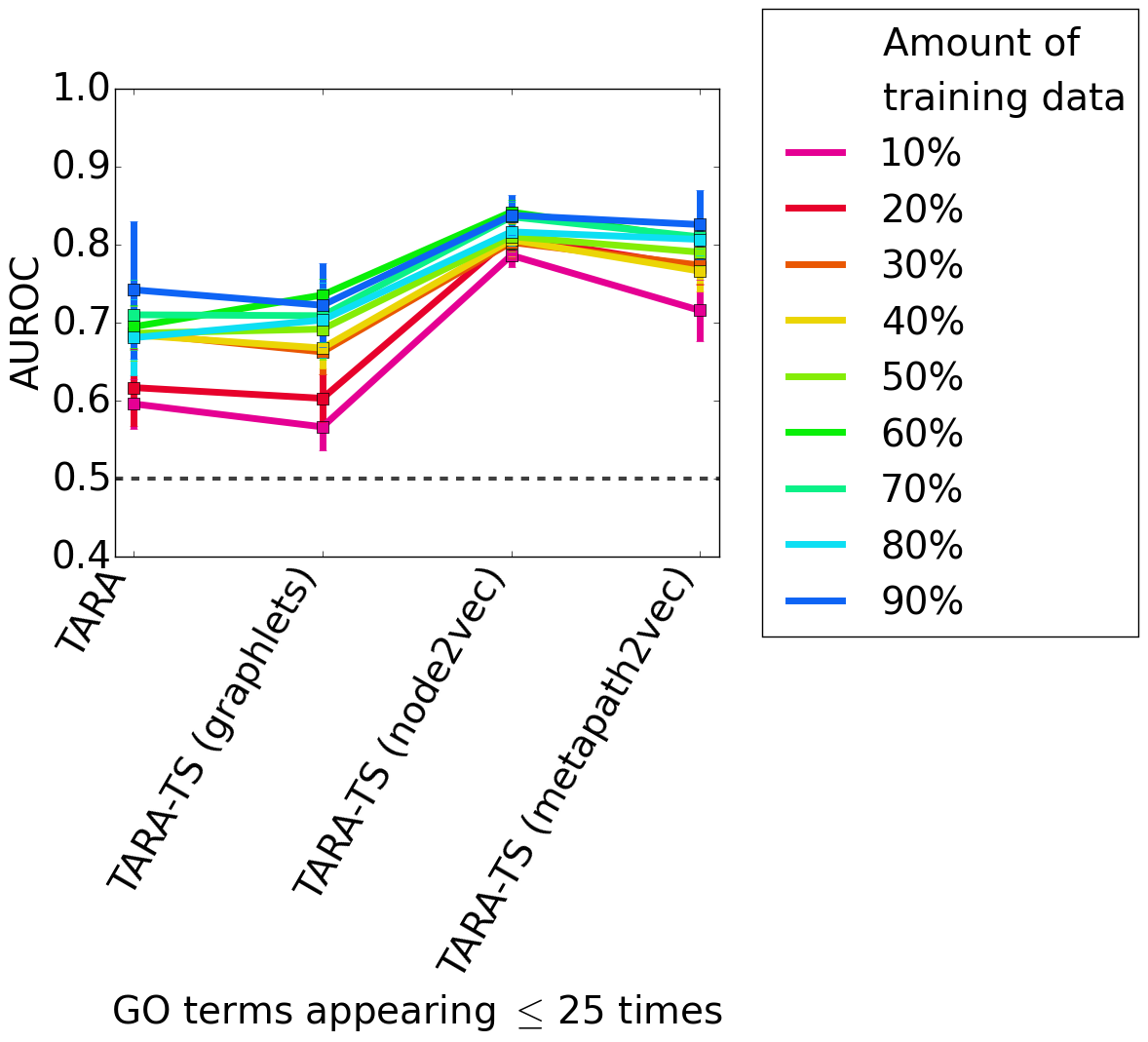}
        \subcaption{}
    \end{subfigure}

    \begin{subfigure}[t]{0.44\textwidth}
        \includegraphics[width=\textwidth]{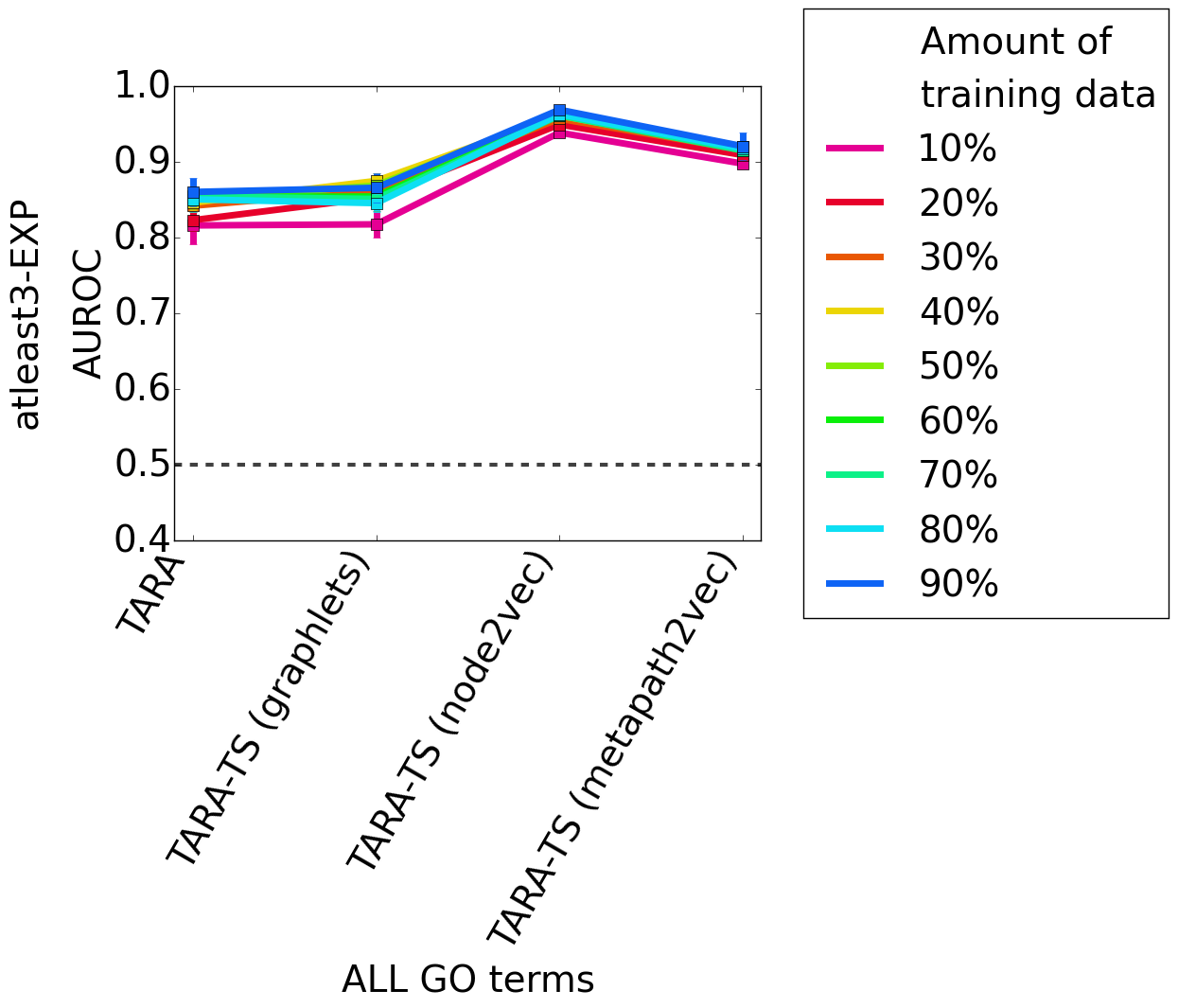}
        \subcaption{}
    \end{subfigure}

 \caption{\label{fig:supp-auroc} \textcolor{black}{Average AUROC of percent training tests for rarity thresholds \textbf{(a, d, g)} ALL, \textbf{(b, e)} 50, and \textbf{(c, f)} 25 using ground truth datasets \textbf{(a, b, c)} atleast1-EXP, \textbf{(d, e, f)} atleast2-EXP, and \textbf{(g)} atleast3-EXP. A dotted black line indicates the AUROC expected if the classifier makes random predictions.}}
\end{figure}

\begin{figure}[ht!]
    \begin{subfigure}[t]{0.26\textwidth}
        \includegraphics[width=\textwidth]{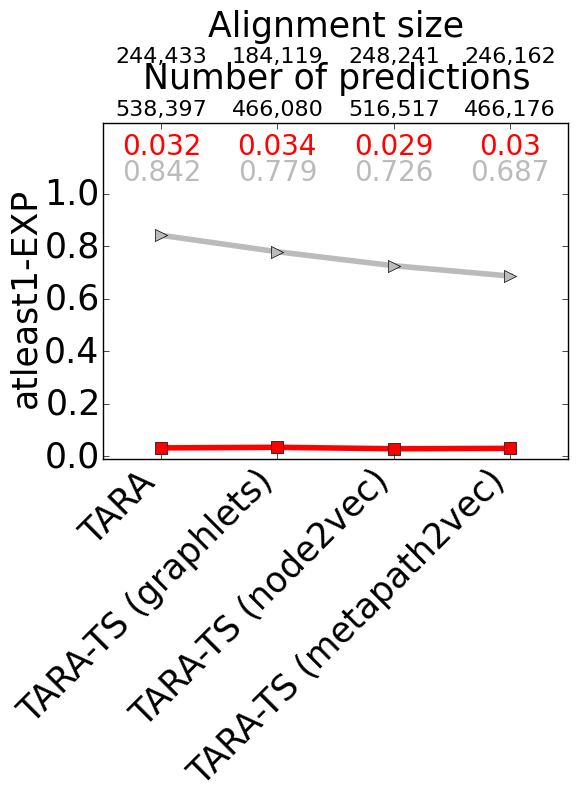}
        \subcaption{}
    \end{subfigure}
    \begin{subfigure}[t]{0.26\textwidth}
        \includegraphics[width=\textwidth]{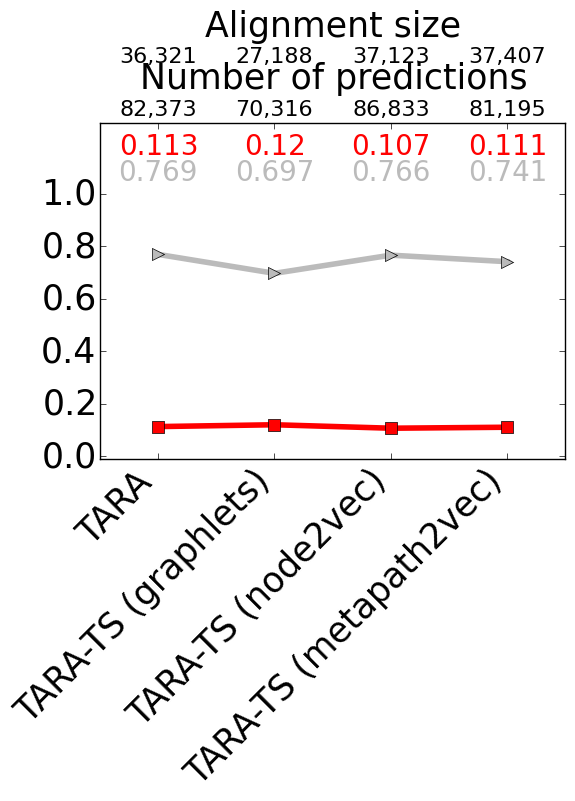}
        \subcaption{}
    \end{subfigure}
    \begin{subfigure}[t]{0.4\textwidth}
        \includegraphics[width=\textwidth]{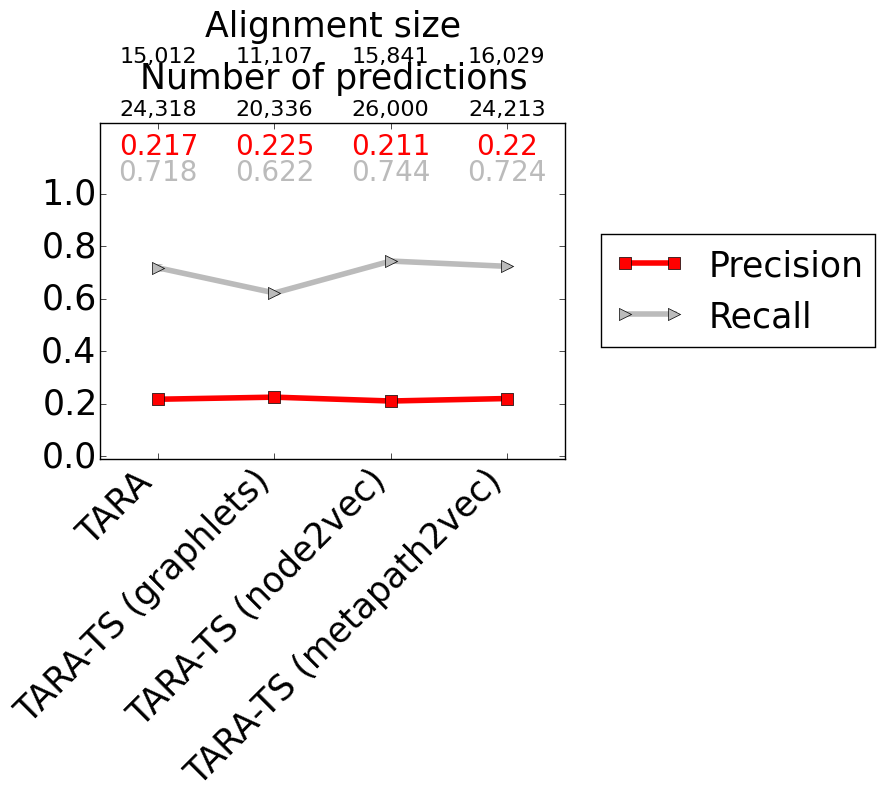}
        \subcaption{}
    \end{subfigure} \\
    
    \begin{subfigure}[t]{0.25\textwidth}
        \includegraphics[width=\textwidth]{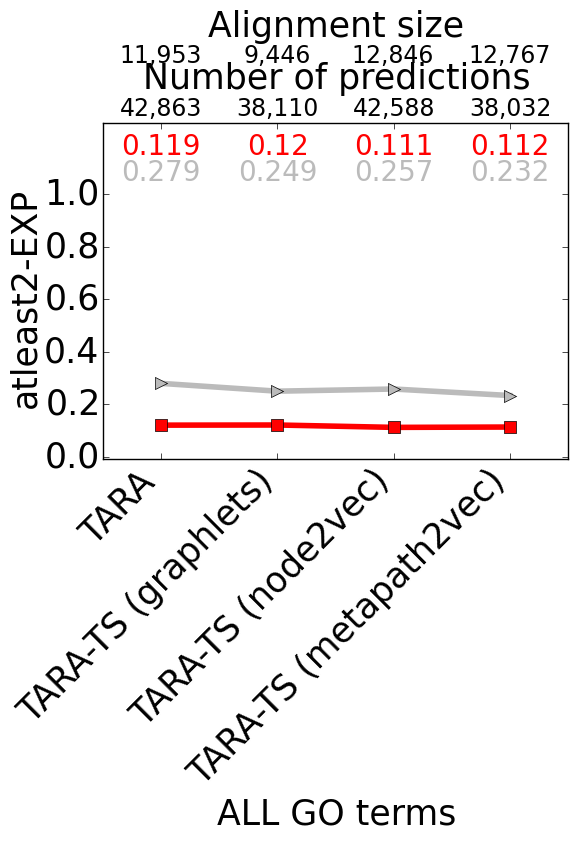}
        \subcaption{}
    \end{subfigure}
    \begin{subfigure}[t]{0.27\textwidth}
        \includegraphics[width=\textwidth]{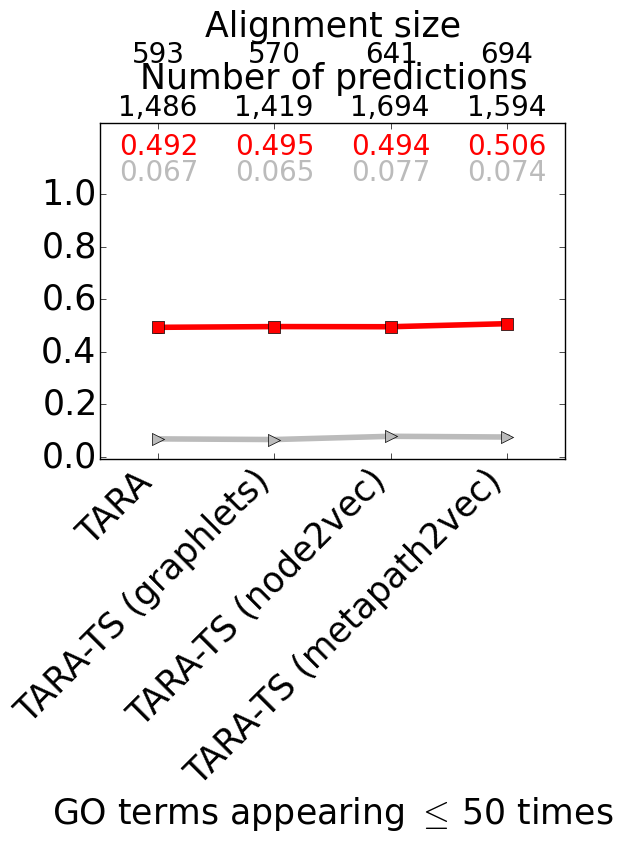}
        \subcaption{}
    \end{subfigure}
    \begin{subfigure}[t]{0.39\textwidth}
        \includegraphics[width=\textwidth]{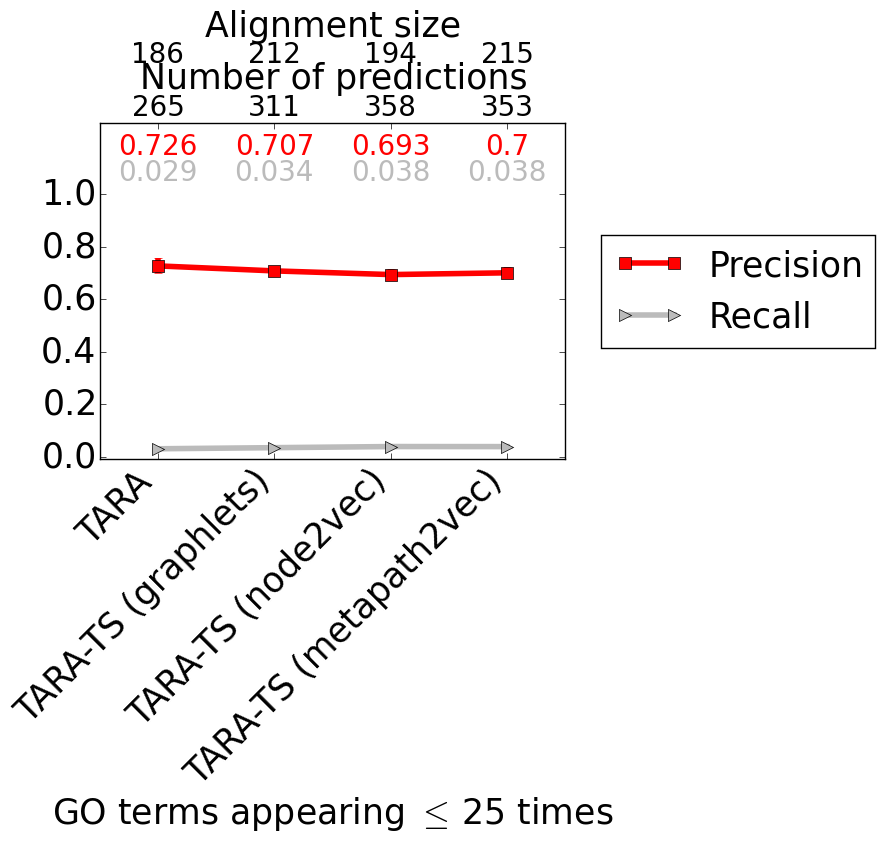}
        \subcaption{}
    \end{subfigure} \\
    
    \begin{subfigure}[t]{0.25\textwidth}
        \includegraphics[width=\textwidth]{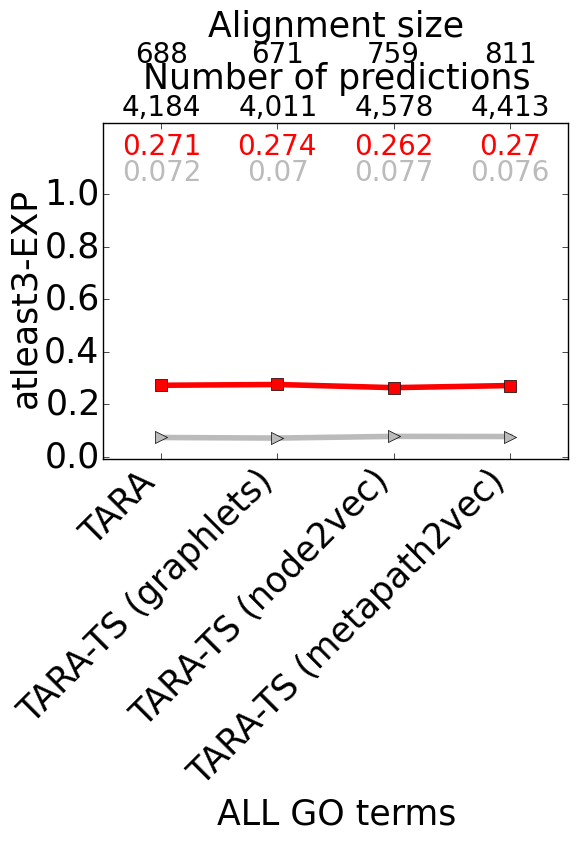}
        \subcaption{}
    \end{subfigure}

 \caption{\label{fig:tara-ts-10} \small{\textcolor{black}{Comparison of TARA and TARA-TS using 10\% of the data as training for rarity thresholds \textbf{(a, d, g)} ALL, \textbf{(b, e)} 50, and \textbf{(c, f)} 25 using ground truth datasets \textbf{(a, b, c)} atleast1-EXP, \textbf{(d, e, f)} atleast2-EXP, and \textbf{(g)} atleast3-EXP in the task of protein functional prediction. The alignment size (i.e., the number of aligned yeast-protein pairs) and number of functional predictions (i.e., predicted protein-GO term associations) made by each method are shown above. For example, the alignment for TARA-10 in \textbf{(a)} contains 244,433 aligned yeast-human protein pairs, and predicts 538,397 protein-GO term associations. Raw precision and recall values are color-coded inside each panel.}}}

\end{figure}

\begin{figure}[ht!]
    \begin{subfigure}[t]{0.26\textwidth}
        \includegraphics[width=\textwidth]{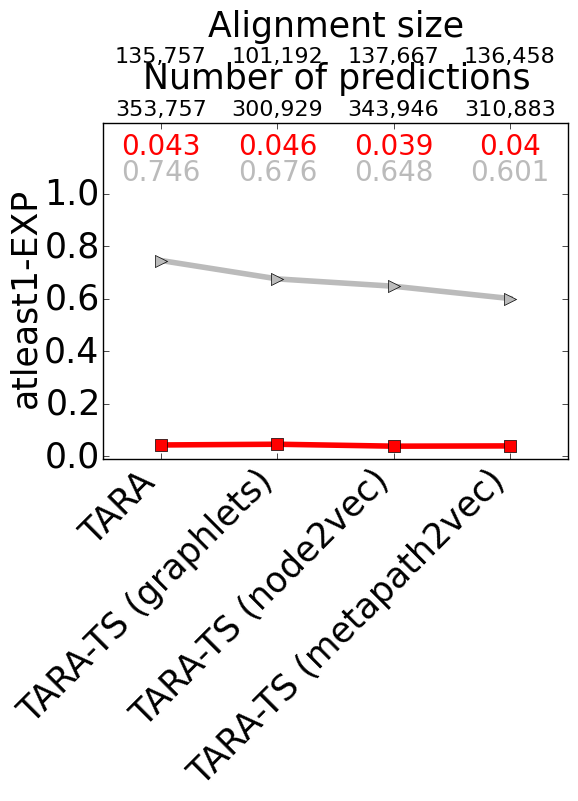}
        \subcaption{}
    \end{subfigure}
    \begin{subfigure}[t]{0.26\textwidth}
        \includegraphics[width=\textwidth]{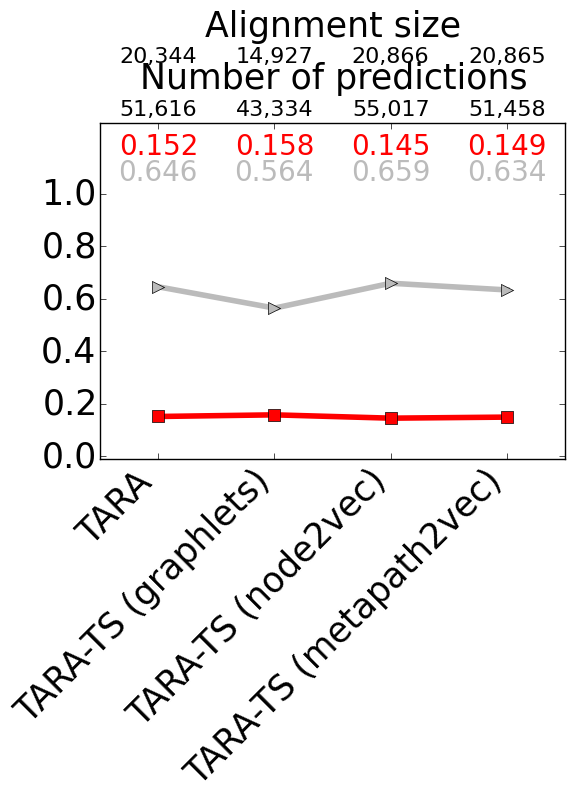}
        \subcaption{}
    \end{subfigure}
    \begin{subfigure}[t]{0.4\textwidth}
        \includegraphics[width=\textwidth]{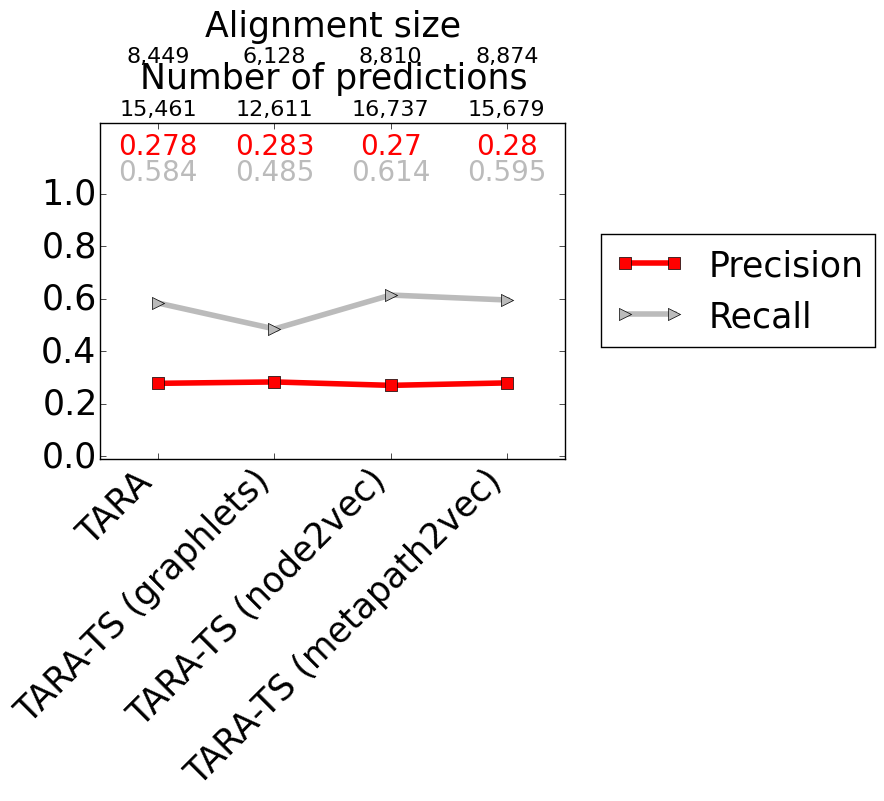}
        \subcaption{}
    \end{subfigure} \\
    
    \begin{subfigure}[t]{0.25\textwidth}
        \includegraphics[width=\textwidth]{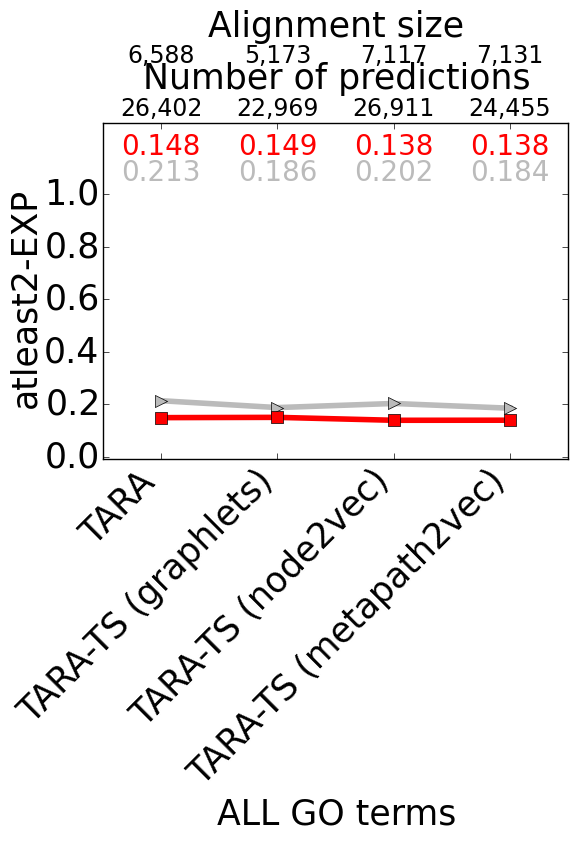}
        \subcaption{}
    \end{subfigure}
    \begin{subfigure}[t]{0.27\textwidth}
        \includegraphics[width=\textwidth]{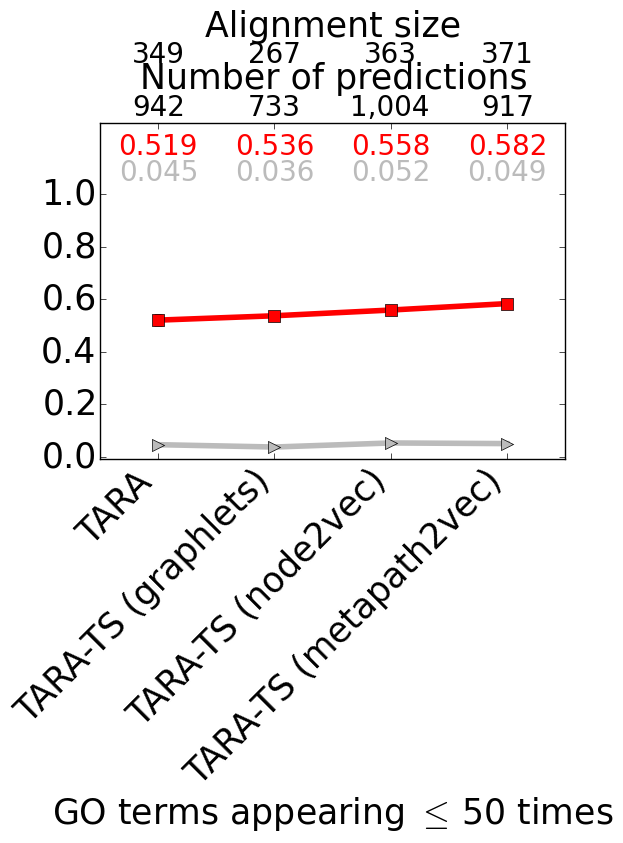}
        \subcaption{}
    \end{subfigure}
    \begin{subfigure}[t]{0.39\textwidth}
        \includegraphics[width=\textwidth]{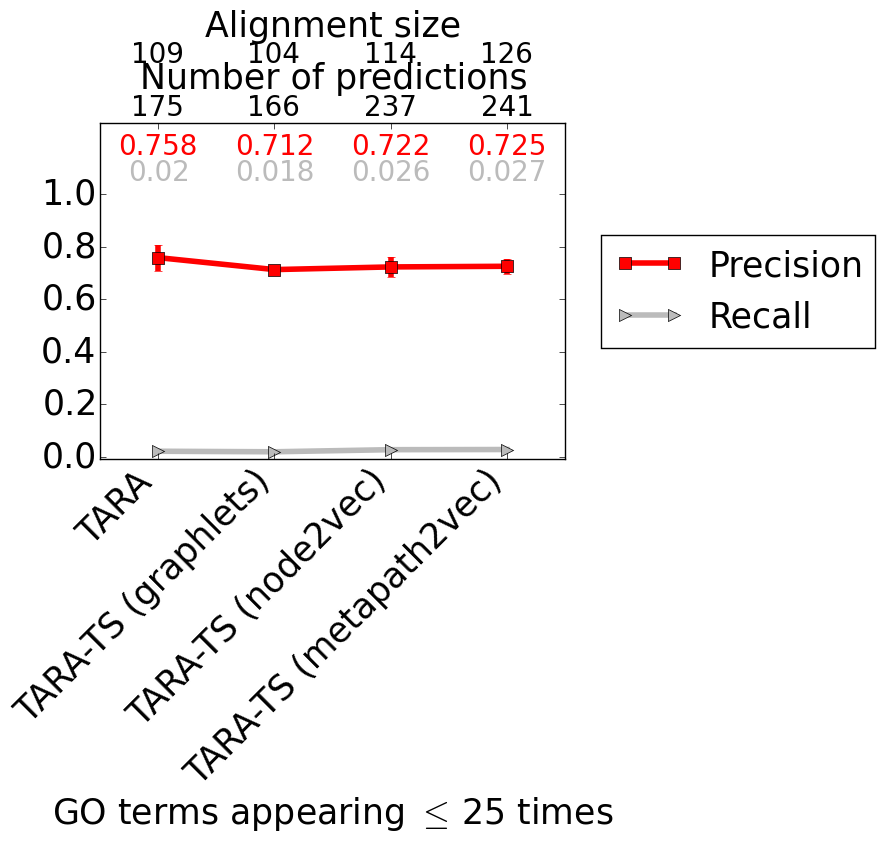}
        \subcaption{}
    \end{subfigure} \\
    
    \begin{subfigure}[t]{0.25\textwidth}
        \includegraphics[width=\textwidth]{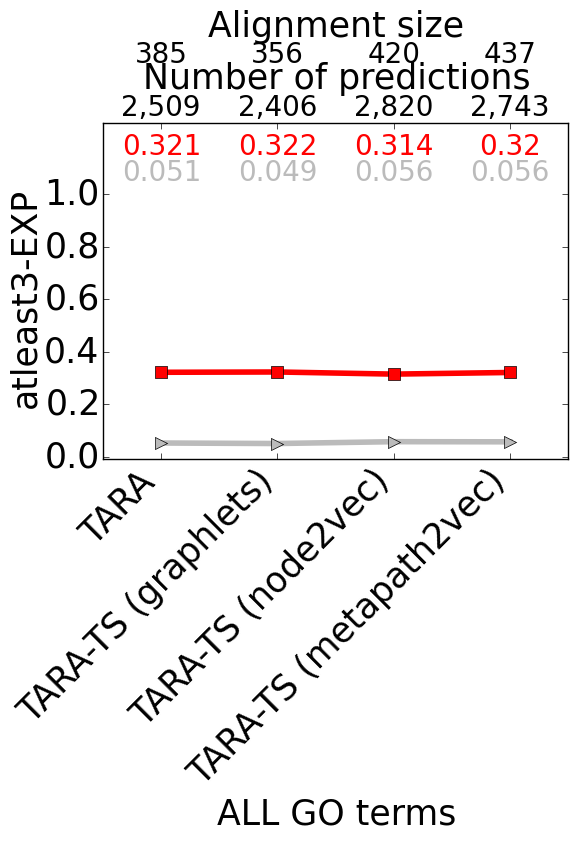}
        \subcaption{}
    \end{subfigure}    

 \caption{\label{fig:tara-ts-50} \small{\textcolor{black}{Comparison of TARA and TARA-TS using 50\% of the data as training for rarity thresholds \textbf{(a, d, g)} ALL, \textbf{(b, e)} 50, and \textbf{(c, f)} 25 using ground truth datasets \textbf{(a, b, c)} atleast1-EXP, \textbf{(d, e, f)} atleast2-EXP, and \textbf{(g)} atleast3-EXP in the task of protein functional prediction. The alignment size (i.e., the number of aligned yeast-protein pairs) and number of functional predictions (i.e., predicted protein-GO term associations) made by each method are shown above. For example, the alignment for TARA-10 in \textbf{(a)} contains 244,433 aligned yeast-human protein pairs, and predicts 538,397 protein-GO term associations. Raw precision and recall values are color-coded inside each panel.}}}
\end{figure}

\begin{figure}[ht!]
    \begin{subfigure}[t]{0.26\textwidth}
        \includegraphics[width=\textwidth]{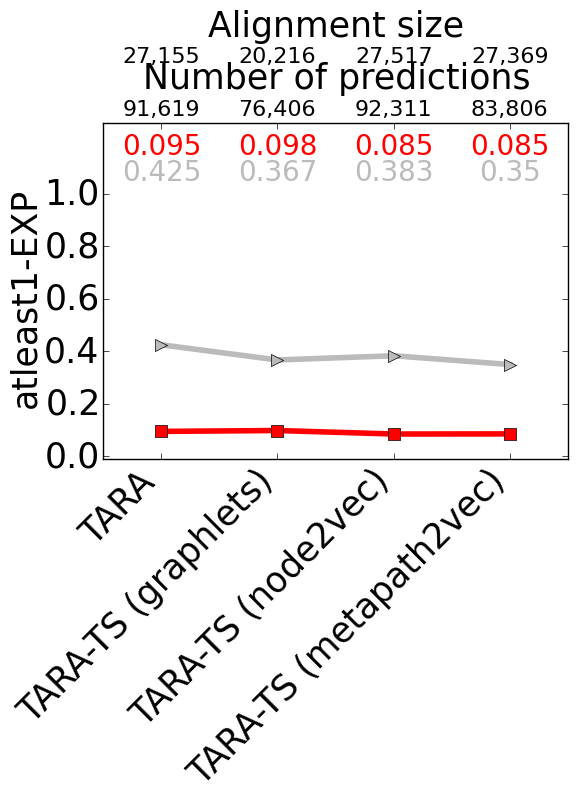}
        \subcaption{}
    \end{subfigure}
    \begin{subfigure}[t]{0.26\textwidth}
        \includegraphics[width=\textwidth]{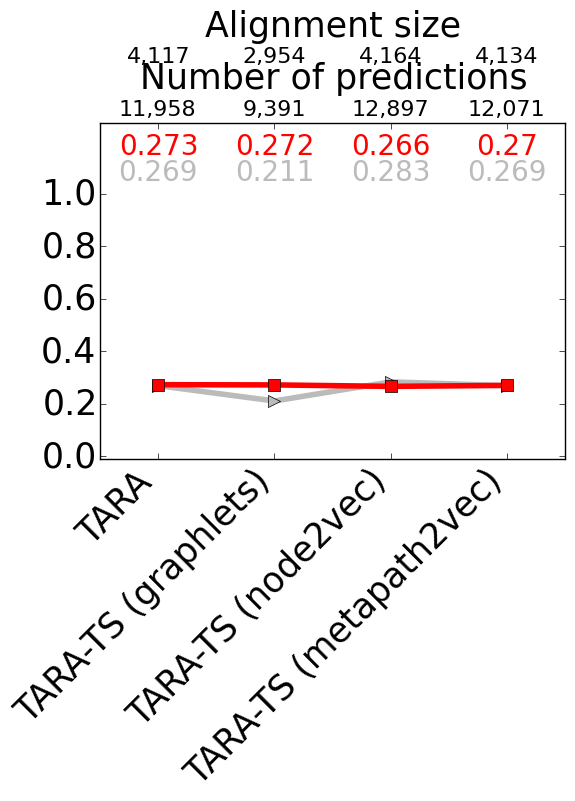}
        \subcaption{}
    \end{subfigure}
    \begin{subfigure}[t]{0.4\textwidth}
        \includegraphics[width=\textwidth]{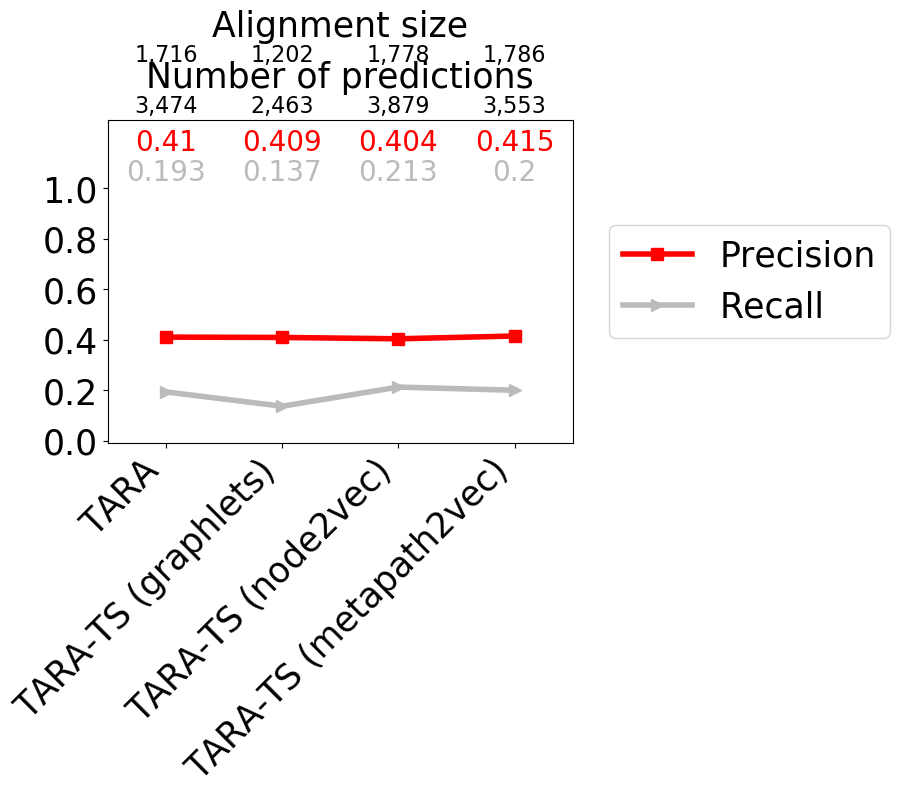}
        \subcaption{}
    \end{subfigure} \\
    
    \begin{subfigure}[t]{0.25\textwidth}
        \includegraphics[width=\textwidth]{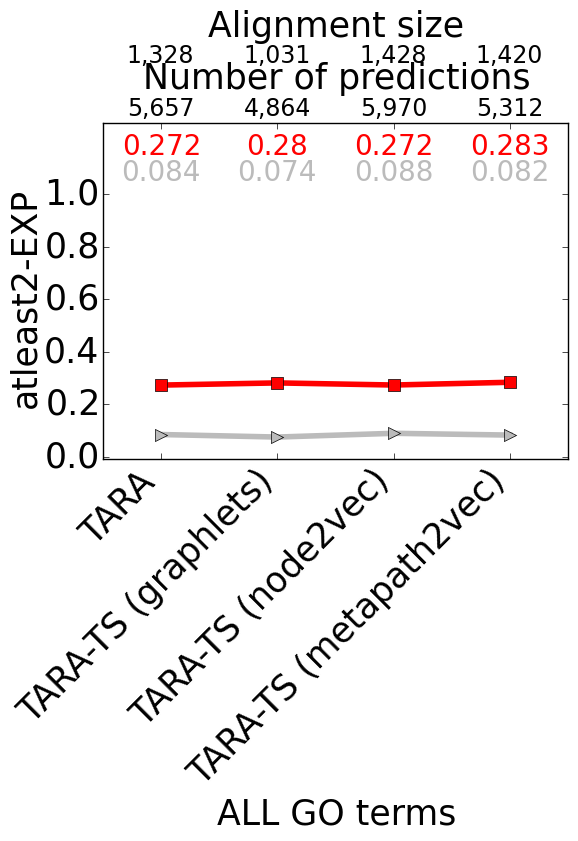}
        \subcaption{}
    \end{subfigure}
    \begin{subfigure}[t]{0.27\textwidth}
        \includegraphics[width=\textwidth]{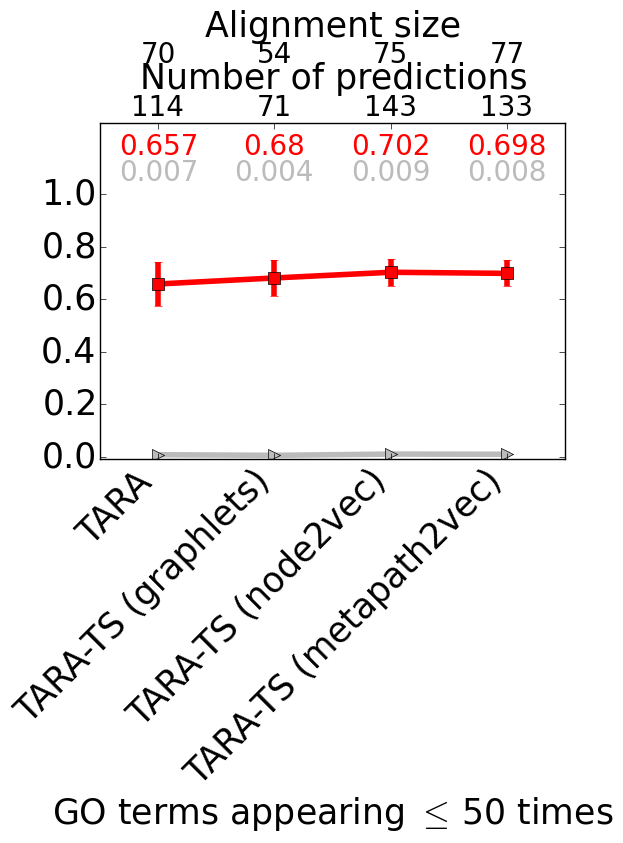}
        \subcaption{}
    \end{subfigure}
    \begin{subfigure}[t]{0.39\textwidth}
        \includegraphics[width=\textwidth]{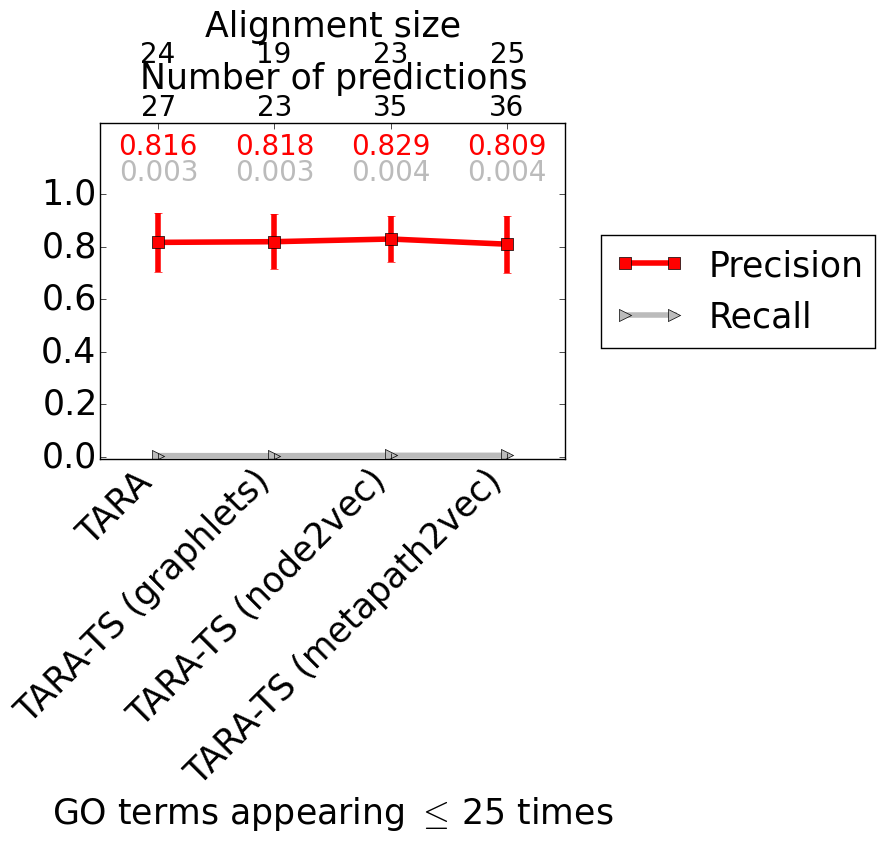}
        \subcaption{}
    \end{subfigure} \\
    
    \begin{subfigure}[t]{0.25\textwidth}
        \includegraphics[width=\textwidth]{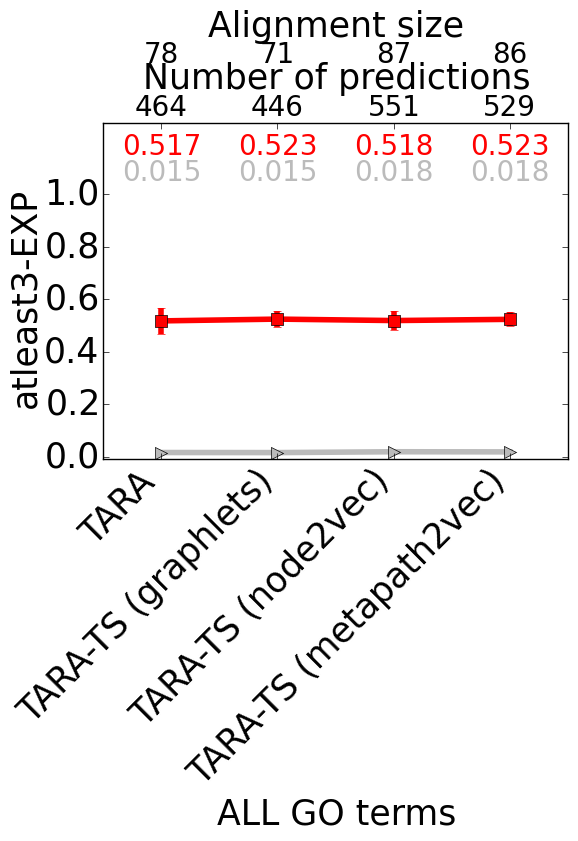}
        \subcaption{}
    \end{subfigure}    

 \caption{\label{fig:tara-ts-90} \small{\textcolor{black}{Comparison of TARA and TARA-TS using 90\% of the data as training for rarity thresholds \textbf{(a, d, g)} ALL, \textbf{(b, e)} 50, and \textbf{(c, f)} 25 using ground truth datasets \textbf{(a, b, c)} atleast1-EXP, \textbf{(d, e, f)} atleast2-EXP, and \textbf{(g)} atleast3-EXP in the task of protein functional prediction. The alignment size (i.e., the number of aligned yeast-protein pairs) and number of functional predictions (i.e., predicted protein-GO term associations) made by each method are shown above. For example, the alignment for TARA-10 in \textbf{(a)} contains 244,433 aligned yeast-human protein pairs, and predicts 538,397 protein-GO term associations. Raw precision and recall values are color-coded inside each panel.}}}
\end{figure}

\begin{figure}[ht!]
    \begin{subfigure}[t]{0.33\textwidth}
        \includegraphics[width=\textwidth]{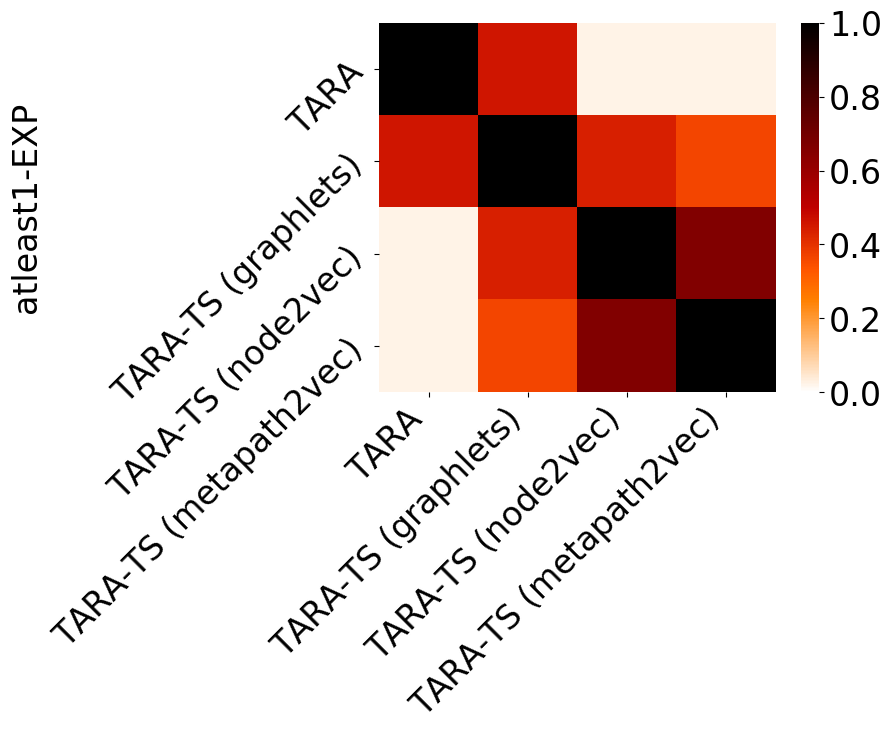}
        \subcaption{}
    \end{subfigure}
    \begin{subfigure}[t]{0.32\textwidth}
        \includegraphics[width=\textwidth]{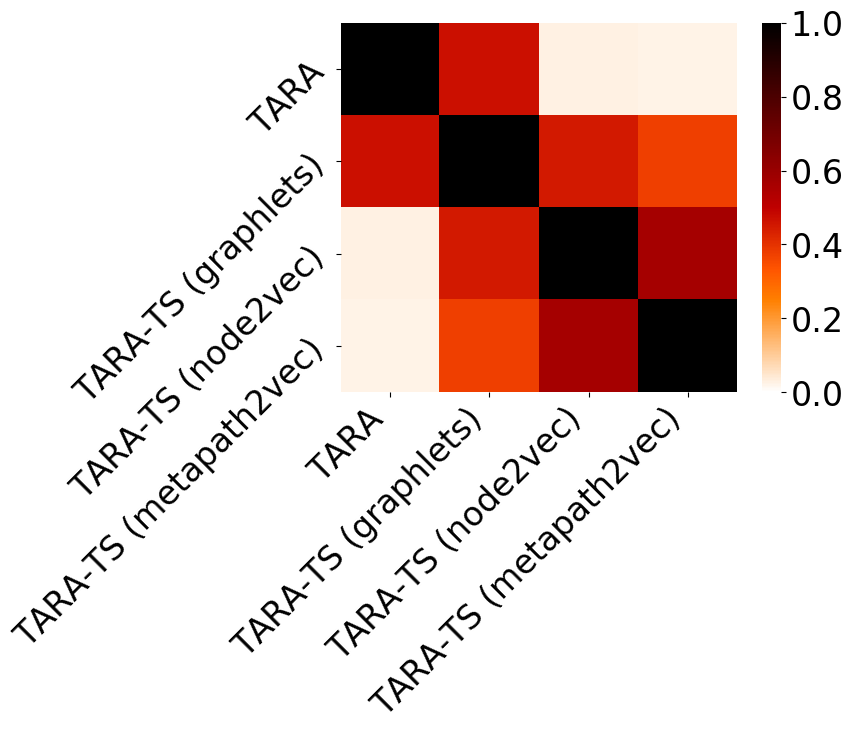}
        \subcaption{}
    \end{subfigure}
    \begin{subfigure}[t]{0.32\textwidth}
        \includegraphics[width=\textwidth]{pairwise-overlaps/perc-overlap-atleast1-EXP-25aln.png}
        \subcaption{}
    \end{subfigure} \\
       \begin{subfigure}[t]{0.33\textwidth}
        \includegraphics[width=\textwidth]{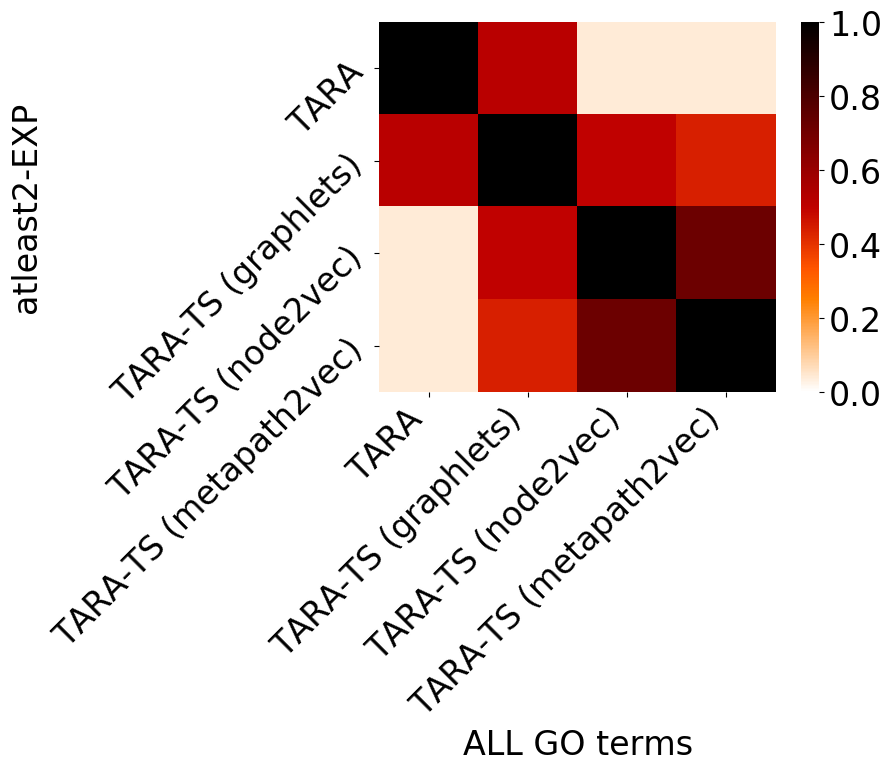}
        \subcaption{}
    \end{subfigure}
    \begin{subfigure}[t]{0.32\textwidth}
        \includegraphics[width=\textwidth]{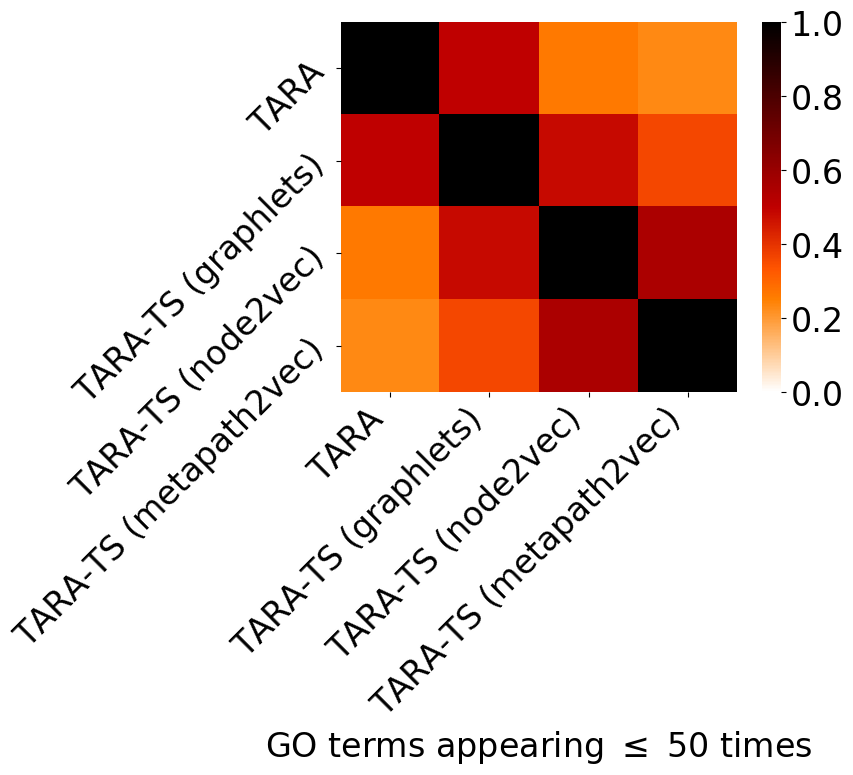}
        \subcaption{}
    \end{subfigure}
    \begin{subfigure}[t]{0.32\textwidth}
        \includegraphics[width=\textwidth]{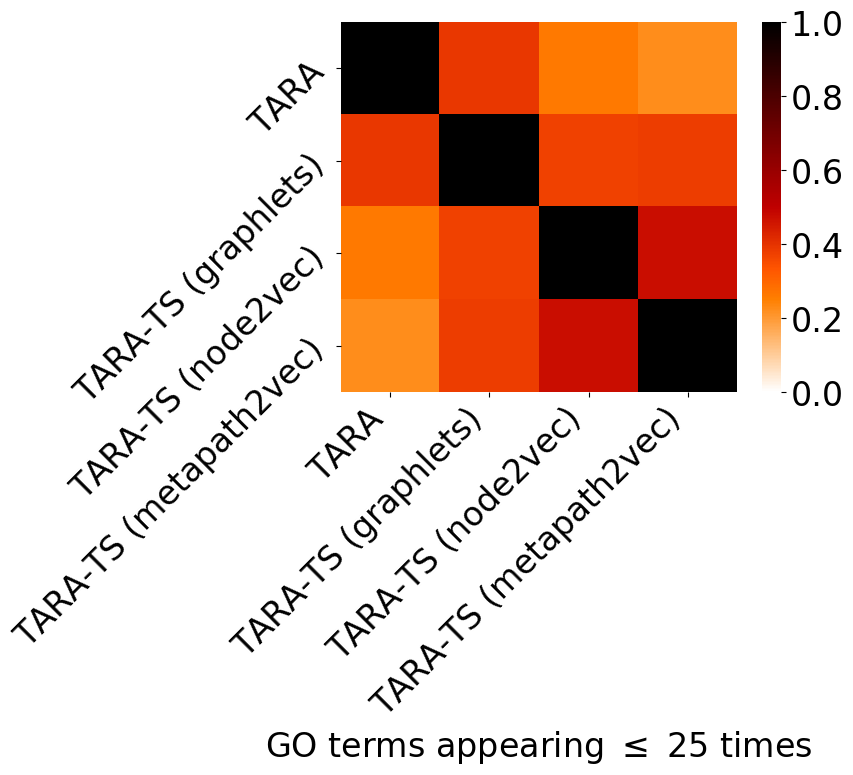}
        \subcaption{}
    \end{subfigure} \\
    
    \begin{subfigure}[t]{0.33\textwidth}
        \includegraphics[width=\textwidth]{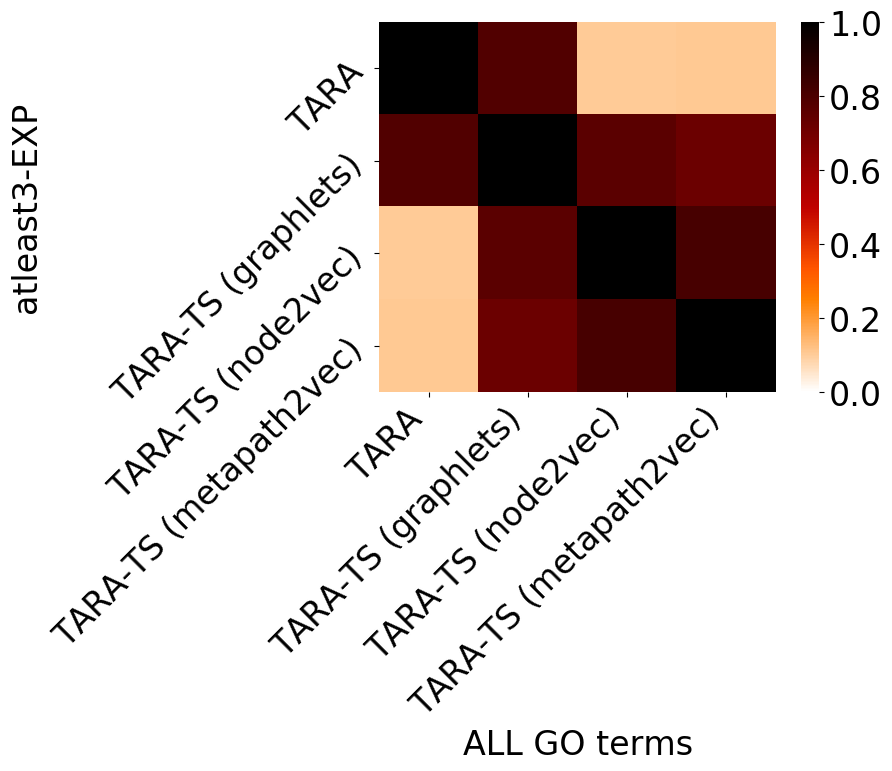}
        \subcaption{}
    \end{subfigure}    

 \caption{\label{fig:tara-ts-alnoverlap} \textcolor{black}{Pairwise overlap, measured by Jaccard index, of the alignments made by TARA and TARA-TS for rarity thresholds \textbf{(a, d, g)} ALL, \textbf{(b, e)} 50, and \textbf{(c, f)} 25 using ground truth datasets \textbf{(a, b, c)} atleast1-EXP, \textbf{(d, e, f)} atleast2-EXP, and \textbf{(g)} atleast3-EXP, using percent training amounts described in Section 3.2.}}
\end{figure}

\begin{figure}[ht!]
    \begin{subfigure}[t]{0.33\textwidth}
        \includegraphics[width=\textwidth]{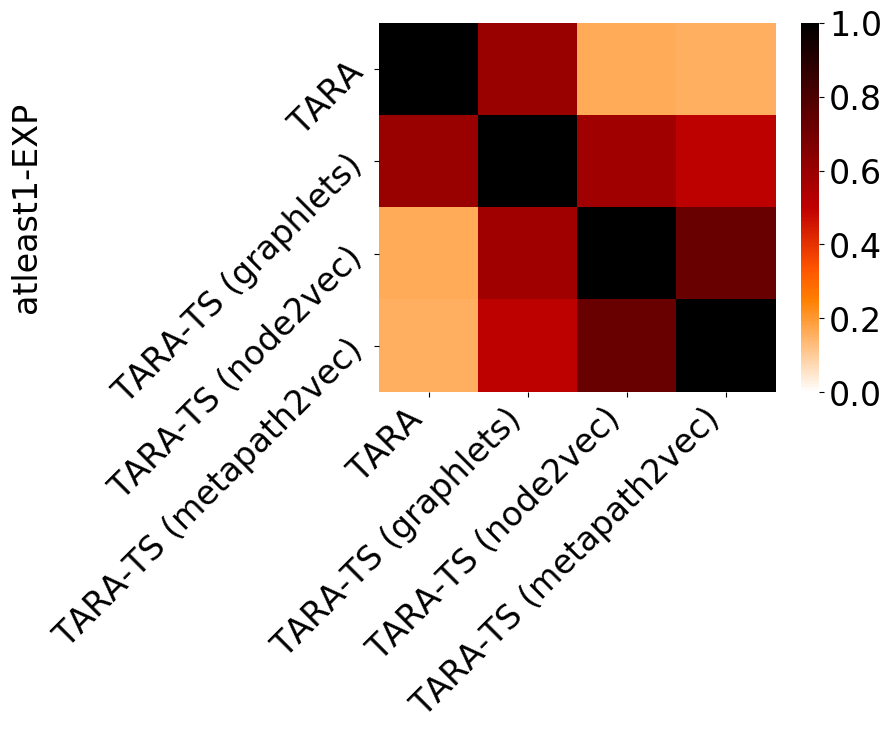}
        \subcaption{}
    \end{subfigure}
    \begin{subfigure}[t]{0.32\textwidth}
        \includegraphics[width=\textwidth]{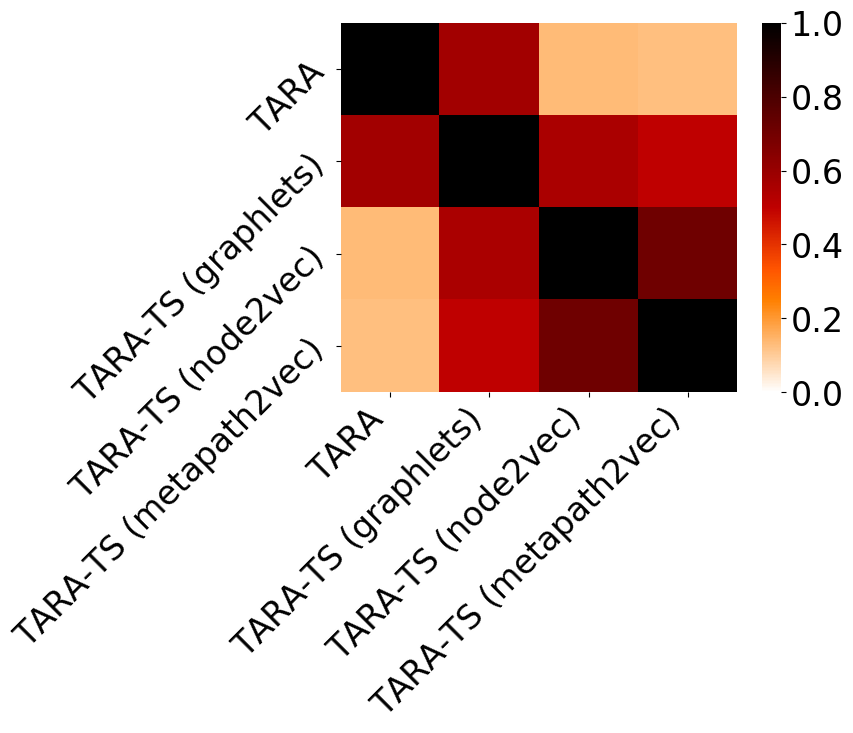}
        \subcaption{}
    \end{subfigure}
    \begin{subfigure}[t]{0.32\textwidth}
        \includegraphics[width=\textwidth]{pairwise-overlaps/perc-overlap-atleast1-EXP-25pred.png}
        \subcaption{}
    \end{subfigure} \\
       \begin{subfigure}[t]{0.33\textwidth}
        \includegraphics[width=\textwidth]{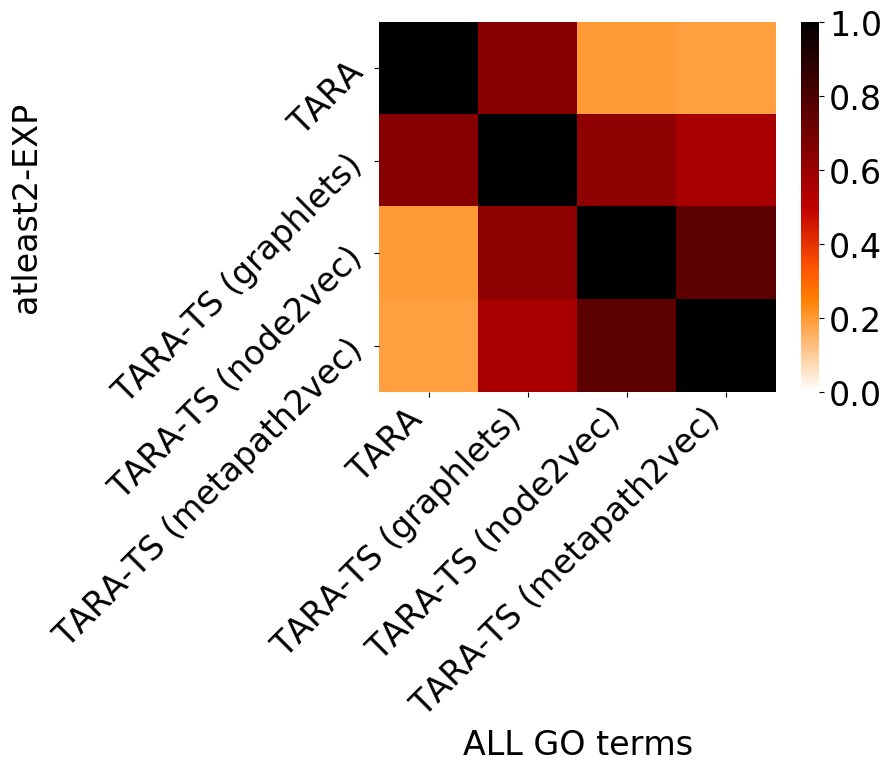}
        \subcaption{}
    \end{subfigure}
    \begin{subfigure}[t]{0.32\textwidth}
        \includegraphics[width=\textwidth]{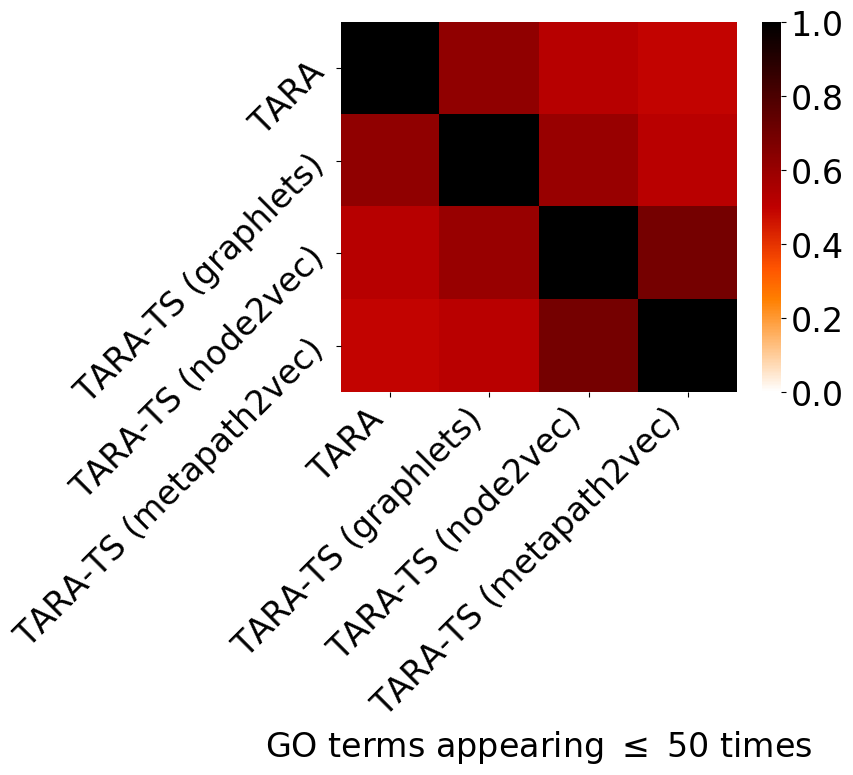}
        \subcaption{}
    \end{subfigure}
    \begin{subfigure}[t]{0.32\textwidth}
        \includegraphics[width=\textwidth]{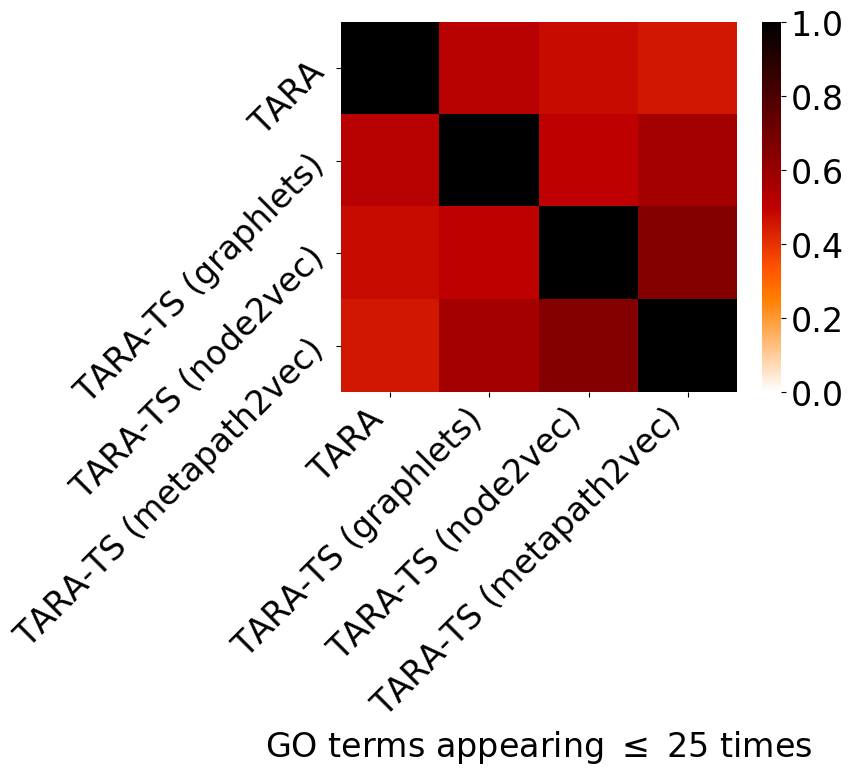}
        \subcaption{}
    \end{subfigure} \\
    
    \begin{subfigure}[t]{0.33\textwidth}
        \includegraphics[width=\textwidth]{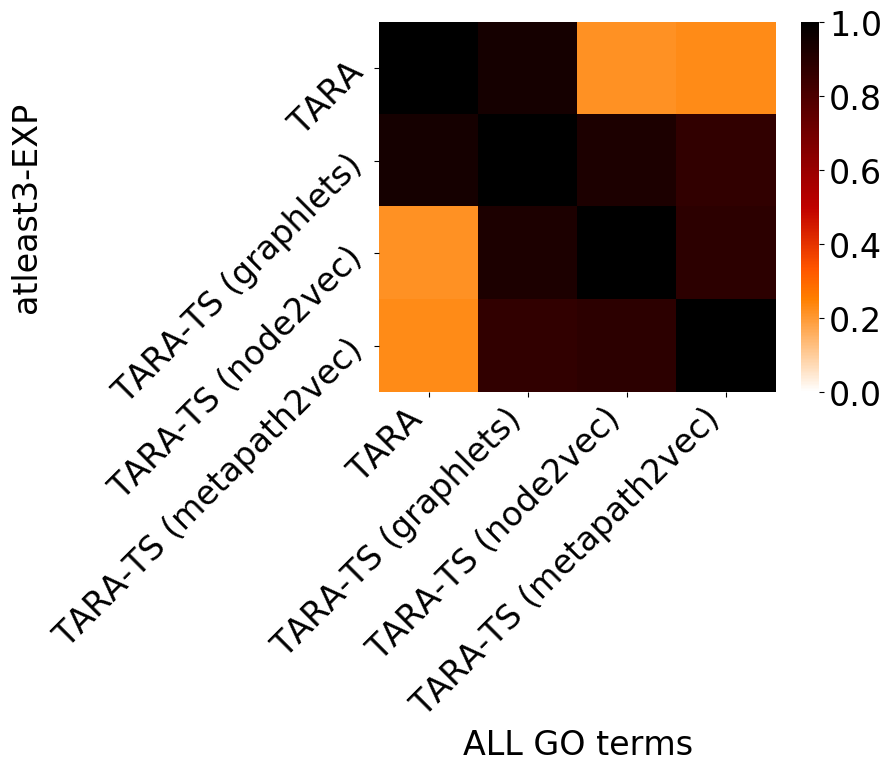}
        \subcaption{}
    \end{subfigure}    

 \caption{\label{fig:tara-ts-predoverlap} \textcolor{black}{Pairwise overlap, measure by Jaccard index, of the predictions made by TARA and TARA-TS for rarity thresholds \textbf{(a, d, g)} ALL, \textbf{(b, e)} 50, and \textbf{(c, f)} 25 using ground truth datasets \textbf{(a, b, c)} atleast1-EXP, \textbf{(d, e, f)} atleast2-EXP, and \textbf{(g)} atleast3-EXP, using percent training amounts described in Section 3.2.}}
\end{figure}

\begin{figure}[ht!]
  
        \begin{subfigure}[t]{0.32\textwidth}
        \includegraphics[width=\textwidth]{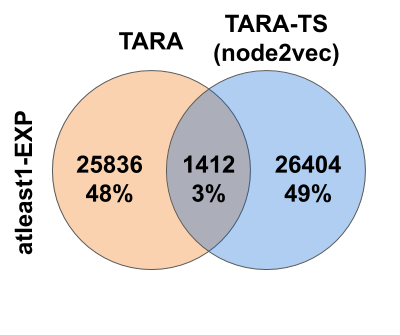}
        \subcaption{}
    \end{subfigure}
    \begin{subfigure}[t]{0.33\textwidth}
        \includegraphics[width=\textwidth]{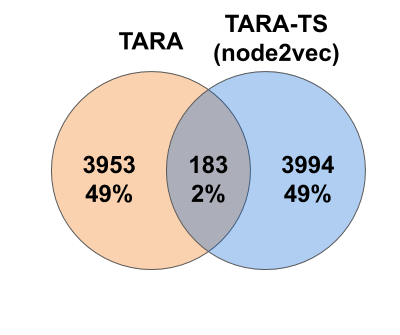}
        \subcaption{}
    \end{subfigure}
    \begin{subfigure}[t]{0.33\textwidth}
        \includegraphics[width=\textwidth]{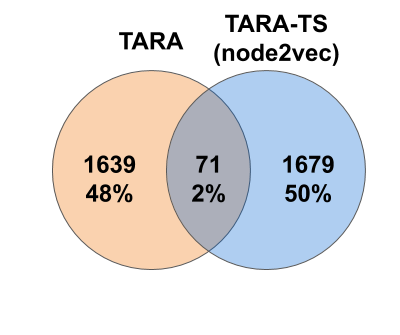}
        \subcaption{}
    \end{subfigure} \\
    
    
    \begin{subfigure}[t]{0.32\textwidth}
        \includegraphics[width=\textwidth]{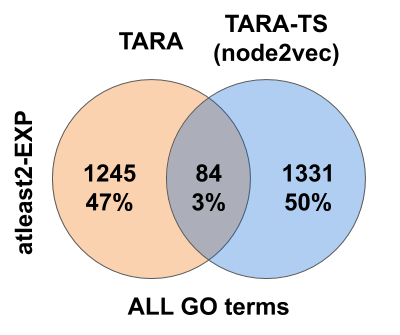}
        \subcaption{}
    \end{subfigure}
    \begin{subfigure}[t]{0.33\textwidth}
        \includegraphics[width=\textwidth]{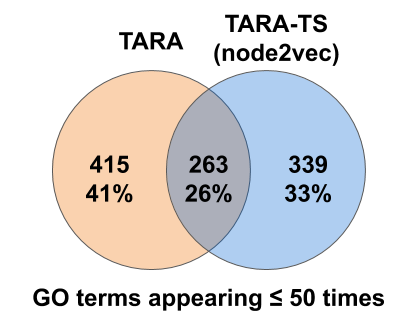}
        \subcaption{}
    \end{subfigure}
    \begin{subfigure}[t]{0.33\textwidth}
        \includegraphics[width=\textwidth]{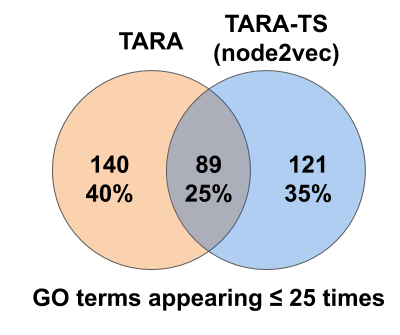}
        \subcaption{}
    \end{subfigure}
    
    \begin{subfigure}[t]{0.32\textwidth}
        \includegraphics[width=\textwidth]{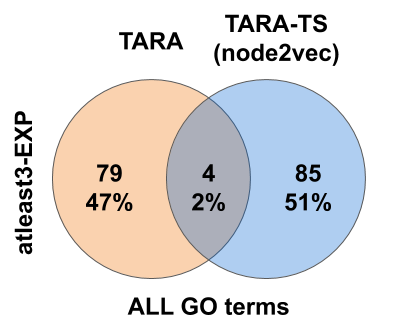}
        \subcaption{}
    \end{subfigure}    

 \caption{\label{fig:aln-overlap} Overlap of the alignments made by TARA and TARA-TS for rarity thresholds \textbf{(a, d, g)} ALL, \textbf{(b, e)} 50, and \textbf{(c, f)} 25 using ground truth datasets \textbf{(a, b, c)} atleast1-EXP, \textbf{(d, e, f)} atleast2-EXP, and \textbf{(g)} atleast3-EXP. Percentages are out of the total number of unique aligned node pairs made by both methods combined. The overlaps are for one of the 10 balanced datasets; so, the alignment size of a method may differ from those in Supplementary Figs. S3-S5, where the statistics are averaged over all balanced datasets.}
\end{figure}

\begin{figure}[ht!]
        \begin{subfigure}[t]{0.32\textwidth}
        \includegraphics[width=\textwidth]{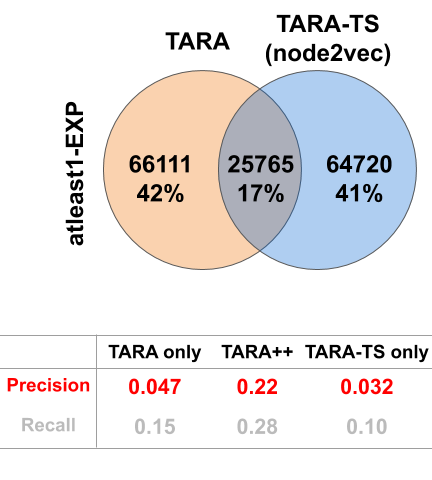}
        \subcaption{}
    \end{subfigure}
    \begin{subfigure}[t]{0.3\textwidth}
        \includegraphics[width=\textwidth]{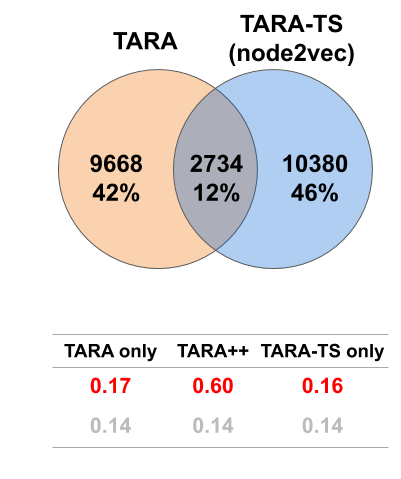}
        \subcaption{}
    \end{subfigure}
    \begin{subfigure}[t]{0.3\textwidth}
        \includegraphics[width=\textwidth]{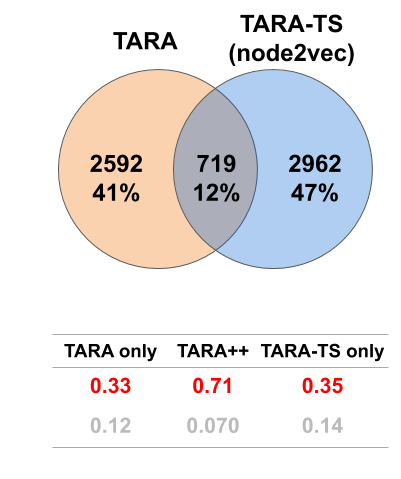}
        \subcaption{}
    \end{subfigure} \\
    
    
    \begin{subfigure}[t]{0.32\textwidth}
        \includegraphics[width=\textwidth]{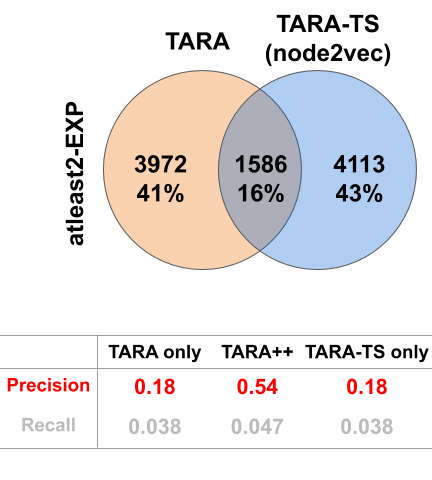}
        \subcaption{}
    \end{subfigure}
    \begin{subfigure}[t]{0.3\textwidth}
        \includegraphics[width=\textwidth]{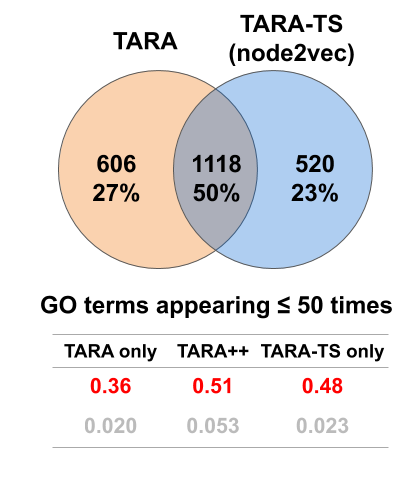}
        \subcaption{}
    \end{subfigure}
    \begin{subfigure}[t]{0.3\textwidth}
        \includegraphics[width=\textwidth]{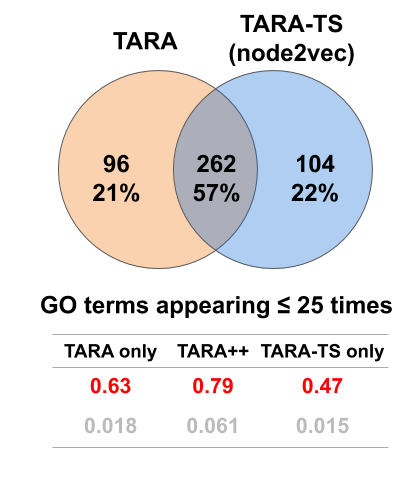}
        \subcaption{}
    \end{subfigure}
    
    \begin{subfigure}[t]{0.32\textwidth}
        \includegraphics[width=\textwidth]{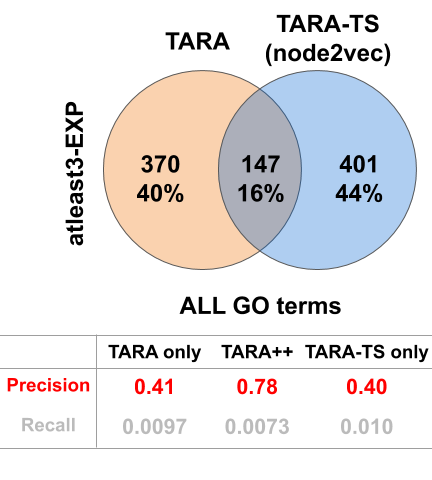}
        \subcaption{}
    \end{subfigure}    

 \caption{\label{fig:pf-pred-overlap} Overlap of the predictions made by TARA and TARA-TS for rarity thresholds \textbf{(a, d, g)} ALL, \textbf{(b, e)} 50, and \textbf{(c, f)} 25 using ground truth datasets \textbf{(a, b, c)} atleast1-EXP, \textbf{(d, e, f)} atleast2-EXP, and \textbf{(g)} atleast3-EXP. Percentages are out of the total number of unique predictions made by both methods combined. Precision and recall are shown for each of the three prediction sets captured by the Venn diagram; TARA++'s predictions are those in the overlap. The overlaps are for one of the 10 balanced datasets; so, the prediction number of a method may differ from those in Supplementary Figs. S3-S5, where the statistics are averaged over all balanced datasets.}
\end{figure}

\begin{figure}[ht!]
    \begin{subfigure}[t]{0.265\textwidth}
        \includegraphics[width=\textwidth]{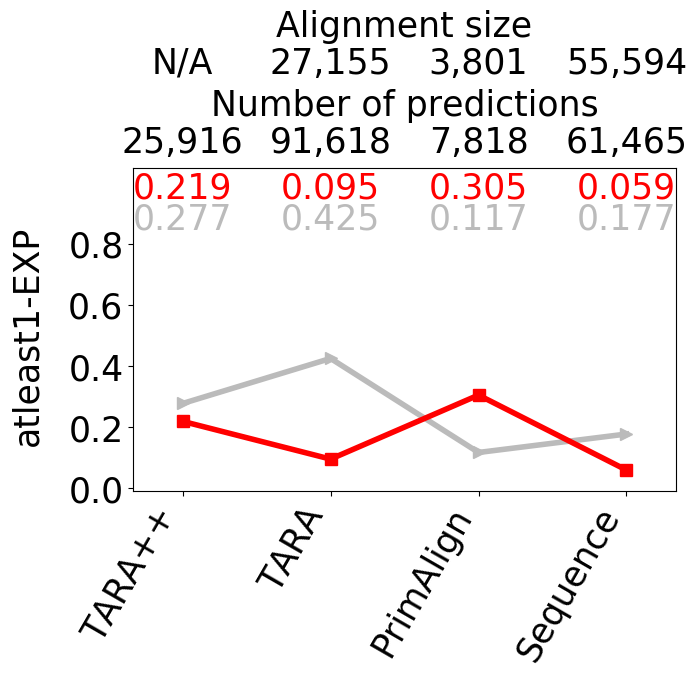}
        \subcaption{}
    \end{subfigure}
    \begin{subfigure}[t]{0.245\textwidth}
        \includegraphics[width=\textwidth]{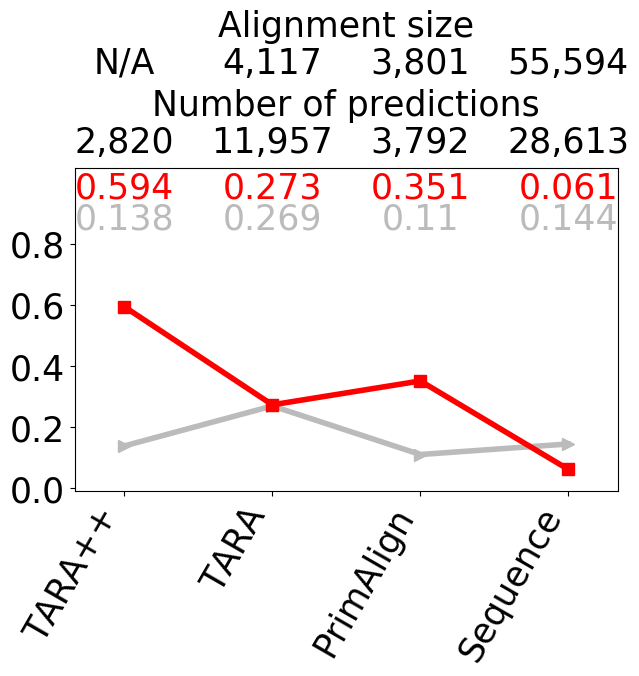}
        \subcaption{}
    \end{subfigure}
    \begin{subfigure}[t]{0.37\textwidth}
        \includegraphics[width=\textwidth]{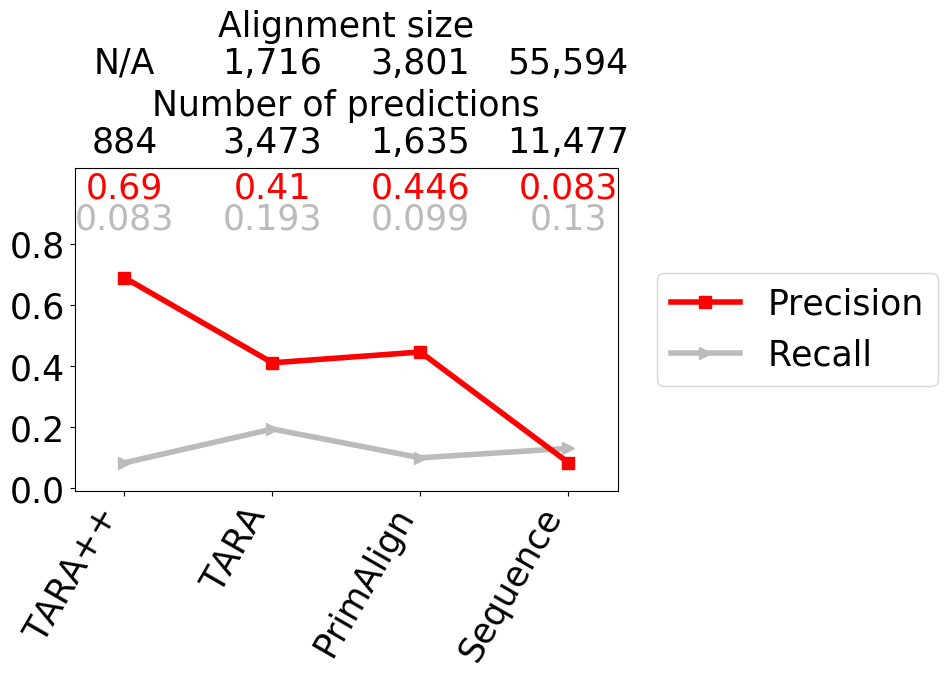}
        \subcaption{}
    \end{subfigure} \\
    
    \vspace{0.1cm}
    
    \begin{subfigure}[t]{0.265\textwidth}
        \includegraphics[width=\textwidth]{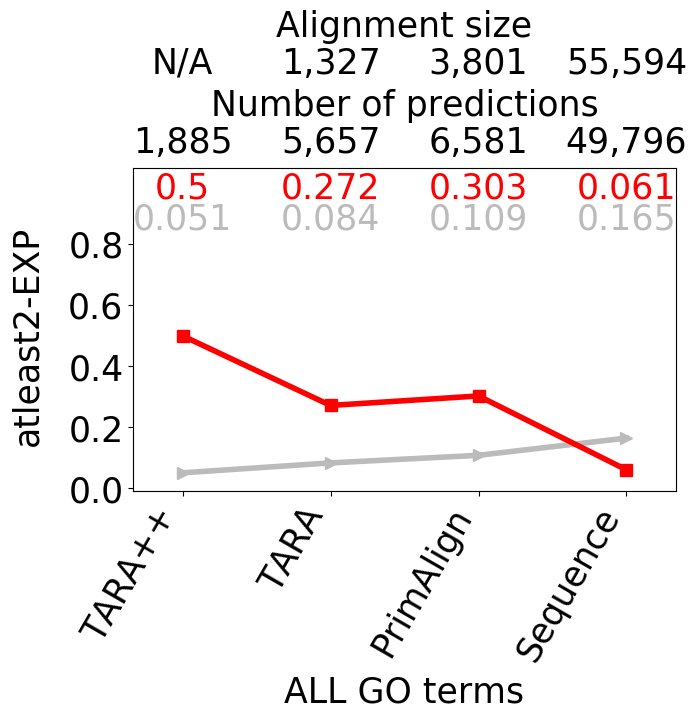}
        \subcaption{}
    \end{subfigure}
    \begin{subfigure}[t]{0.245\textwidth}
        \includegraphics[width=\textwidth]{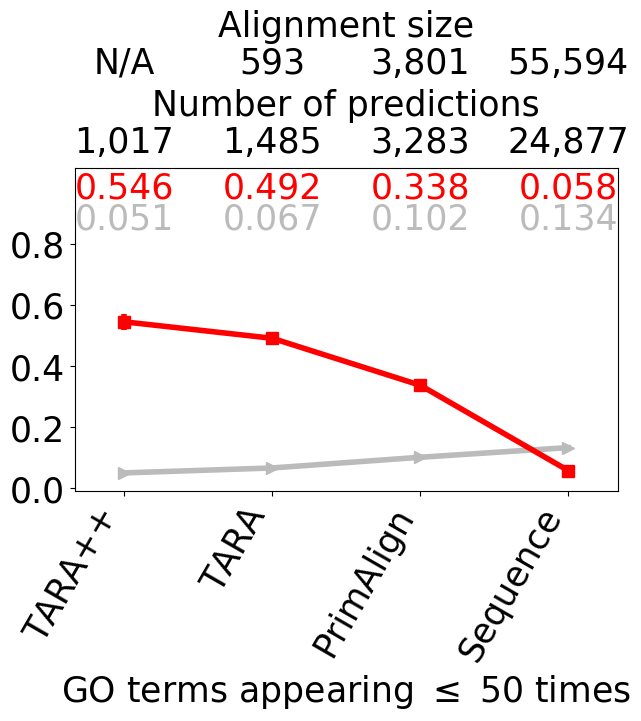}
        \subcaption{}
    \end{subfigure}
    \begin{subfigure}[t]{0.365\textwidth}
        \includegraphics[width=\textwidth]{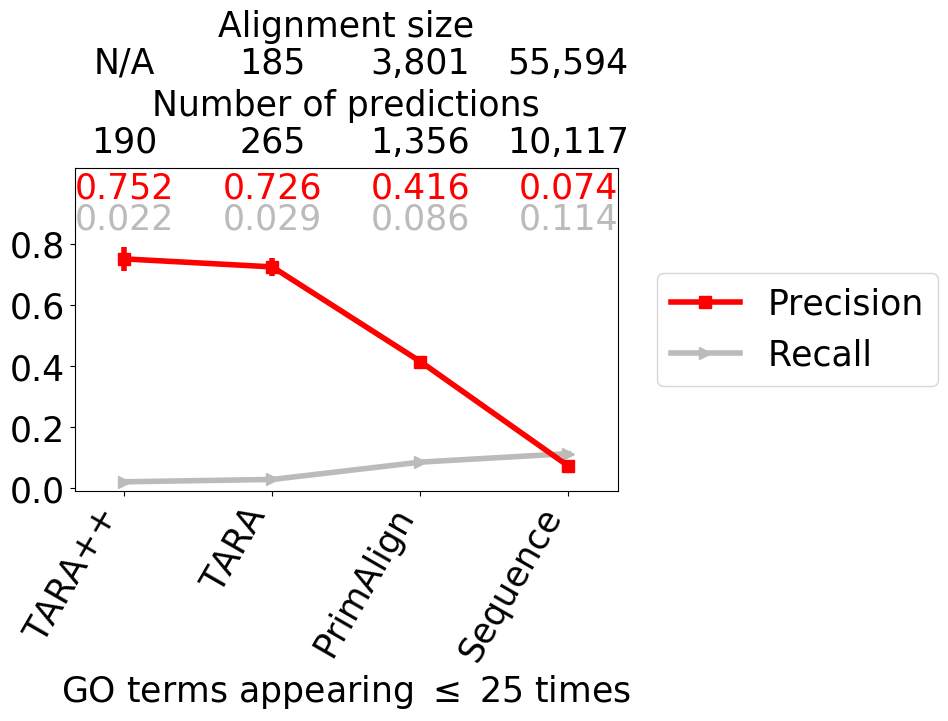}
        \subcaption{}
    \end{subfigure}
    
    \begin{subfigure}[t]{0.265\textwidth}
        \includegraphics[width=\textwidth]{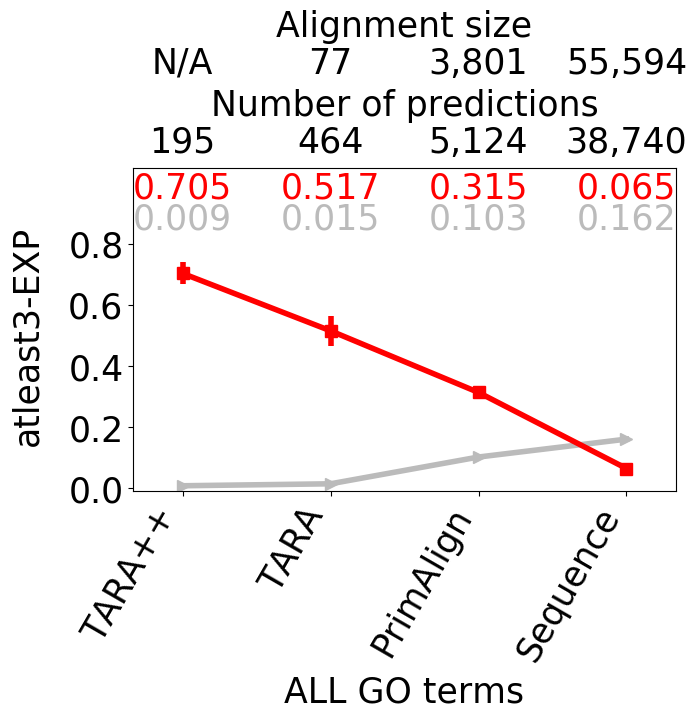}
        \subcaption{}
    \end{subfigure}    
 \caption{\label{fig:tarapp-vs-all} Comparison of four NA methods for rarity thresholds \textbf{(a, d, g)} ALL, \textbf{(b, e)} 50, and \textbf{(c, f)} 25 using ground truth datasets \textbf{(a, b, c)} atleast1-EXP, \textbf{(d, e, f)} atleast2-EXP, and \textbf{(g)} atleast3-EXP in the task of protein functional prediction. The alignment size (i.e., the number of aligned yeast-protein pairs) and number of functional predictions (i.e., predicted protein-GO term associations) made by each method are shown above, except that TARA++ does not have an alignment \emph{per se}. i.e., TARA++ comes from the overlap of \emph{predictions} made by TARA and TARA-TS; hence the ``N/A''s. For example, the alignment for TARA in \textbf{(a)} contains 27,155 aligned yeast-human protein pairs, and predicts 91,618 protein-GO term associations. Raw precision and recall values are color-coded inside each panel. For TARA++ and TARA, results are averages over all balanced datasets; the standard deviations are small and thus invisible.}
\end{figure}

\begin{table}[ht!]
\centering
\begin{tabular}{l | rrr}
\hline
              & atleast1-EXP & atleast2-EXP & atleast3-EXP \\
\hline
TARA-TS & 3811 & 480 & 444 \\
TARA   & 8090 & 4676 & 4634 \\
PrimAlign      & 16        & 16        & 16        \\
Sequence & N/A & N/A & N/A \\
\hline
\end{tabular}
\vspace{-0.2cm}
\caption{\label{fig:supp-aln-times} Running times (in seconds) of TARA-TS, TARA, PrimAlign, and Sequence, when considering ALL GO terms. TARA++'s running time is a function of TARA-TS's and TARA's (see Section 3.3 in the main paper). We use a precomputed alignment for Sequence  (see Section 3.3 in the main paper), hence the ``N/A''s. }
\end{table}